\documentclass[
aps,
prd,
preprint,
amsmath,
amssymb,
floats,
nofootinbib,
tightenlines,
showpacs,
10pt]{revtex4-1}
\usepackage[small, it]{caption}
\usepackage{latexsym} 
\usepackage{amsmath, mathtools}
\usepackage{amssymb, amsthm}
\usepackage{graphicx}
\usepackage{rotating}
\usepackage[english]{babel}
\newcommand{\sla}[1]{/\!\!\!#1}

\newcommand{\GeV}{{\ensuremath\rm GeV}}
\newcommand{\TeV}{{\ensuremath\rm TeV}}

\begin{document}



\title{Distorted Mass Edges at LHC \\ from Supersymmetric Leptoquarks}

\begin{flushright}
DESY 11-082 \\ FR-PHENO-2010-033  \\ EDINBURGH-2010-26
\end{flushright}

\author{J\"urgen Reuter}
 \email{juergen.reuter@desy.de}
 \affiliation{DESY Theory Group, Notkestr. 85, D--22603 Hamburg, Germany}
 \affiliation{University of Edinburgh, School of Physics and
   Astronomy, JCMB, The King's Buildings, Mayfield Road, Edinburgh EH9
   3JZ, SCOTLAND} 
 \affiliation{Albert-Ludwigs-Universit\"at Freiburg, Physikalisches
   Institut, Hermann-Herder-Str. 3, D--79104 Freiburg, GERMANY}

\author{Daniel Wiesler}
 \email{daniel.wiesler@desy.de}
 \affiliation{DESY Theory Group, Notkestr. 85, D--22603 Hamburg, Germany}
 \affiliation{University of Edinburgh, School of Physics and
   Astronomy, JCMB, The King's Buildings, Mayfield Road, Edinburgh EH9
   3JZ, SCOTLAND} 
 \affiliation{Albert-Ludwigs-Universit\"at Freiburg, Physikalisches
   Institut, Hermann-Herder-Str. 3, D--79104 Freiburg, GERMANY}

\date{\today}

\begin{abstract}%
Supersymmetric (SUSY) Grand Unified Theories based on exceptional gauge
groups like $E_6$ have recently triggered a lot of interest. Aside from
top-down motivations, they contain phenomenologically interesting states
with leptoquark quantum numbers. Their SUSY partners, leptoquarkinos, 
will appear like all $R$-odd particles in decay cascades, but mass
edges in kinematic distributions -- originating from the same
semi-exclusive final states -- will however have major differences to
the corresponding edges of ordinary squarks. This distortion of
standard observables bears the opportunity to detect them at LHC, but
may also pose a severe confusion of underlying model assumptions,
which should be handled with care and, if interpreted falsely, might
even prevent a possible discovery.  
\end{abstract}


\maketitle

\section{Introduction}

Supersymmetry is one of the most promising solutions of the
hierarchy and fine-tuning problem, namely the vast difference between
the electroweak (EW) and the Planck scale, and the very stability of
this difference. It yields a mechanism for radiatively generating EW
symmetry breaking, allows for an exact unification of all forces and
conveys a candidate for dark matter. However, it comes with the price
of having new problems, connected to the flavour sector, the
stability of the proton, and new sorts of hierarchy problems known as
the $\mu$ problem and doublet-triplet splitting. To address these
questions, models have been developed that derive from a Planck or
GUT scale exceptional gauge group like
$E_6$~\cite{Hewett:1988xc}, and might be embedded in the context of the
heterotic string. Such $E_6$-based models have a matter-Higgs
unification, are automatically anomaly-free, include the right-handed
neutrino, and solve the $\mu$ problem as an effective next-to-minimal
SUSY Standard Model (NMSSM). However, one either has to solve a
problem similar to doublet-triplet splitting, or use e.g. an
intermediate left-right symmetric model, where the high-scale braking
happens through orbifold compactifications~\cite{Braam:2010sy,Braam:2011}. 
The fundamental representation of $E_6$, the ${\bf 27}$, contains exotic states
which carry both lepton and baryon number and hence act as lepto- or diquarks. 
This is why a mechanism as in \cite{Braam:2010sy} is chosen specifically to 
prevent a phenomenologically disastrous rapid proton decay. 
As these exotics are left-chiral superfields (with vector-like quantum numbers
with respect to the EW gauge group), they come as a pair of scalars, 
$D$ and $D^*$, being $R$ even, and a Dirac fermion, $\tilde{D}$, being
$R$ odd, at the EW scale. The states are called leptoquarks and
leptoquarkinos, respectively. Their potential discovery at the Large
Hadron Collider (LHC) may allow for a direct handle on the GUT structure
of these models at the TeV scale beyond super-precise extrapolation
of parameters over 13-15 orders of magnitude. 

For the rest of this letter, it is sufficient to take the model-building set-up
above as a rough motivation how such states could come about in
Nature, and further on just assume their existence together with the
spectrum of an NMSSM-like model. The phenomenology of the scalar
leptoquarks is very similar to that of non-supersymmetric
states due to their R-even nature and will be discussed in a following
publication~\cite{Braam:2011}. While the pair 
production of the fermionic superpartners, the
leptoquarkinos, is almost completely determined by QCD, their decays as
$R$-odd particles show the very same cascade-like structures as
squark and gluino decays. However, their decay products contain both
non-vanishing lepton and baryon number. Hence, kinematic edge
structures for the mass determination of new physics states derived
from jet-lepton or jet-dilepton exclusive final states have very
characteristic features which -- using invalid assumptions about the
underlying SUSY model -- could lead to wrong particle identifications
and mass determinations (The latter point is particularly relevant, if
the scalar states which happen to be usually heavier than the fermions
might lie outside the kinematic reach of LHC). The intent of this
paper is to first introduce the relevant observables before a brief
discussion of the exotic fermion production and decay mechanisms is
followed by a dedicated study on the impact of leptoquarkinos with
varying masses onto SUSY squark analysis methods. Here it is the goal
to show the essential and important differences between standard SUSY
squark cascades and leptoquarkino-triggered cascades, and why it is
important to consider at least both, if not all possible model
alternatives.


\section{LHC mass edges from leptoquarkinos}

Mass edge variables
\cite{Lester:2001zx,Bachacou:1999zb,Lester:2007fq,Brooijmans:2010tn,Barr:2010zj}
have mostly been developed with a certain decay pattern in mind:
left-handed squark into a quark, two leptons and the lightest
neutralino via the on-shell decays of the second-to-lightest
neutralino and a right-handed slepton:  

\begin{eqnarray}
\tilde{q}_L \rightarrow q \tilde{\chi}^0_2 \rightarrow
ql^\pm\tilde{l}^\mp_{R}  \rightarrow q l^\pm l^\mp \tilde{\chi}^0_1 
\end{eqnarray}

Since one is not able to distinguish experimentally which of the
leptons $l^\pm$ and $l^\mp$ is nearest to the quark (in terms of the decay
  cascade), two specific observables have been introduced
\cite{Bachacou:1999zb} to allow for a discrimination: 
\begin{align}
  m_{ql,high} = \max \{m_{ql^+},m_{ql^-}\}\\ 
  m_{ql,low} = \min \{m_{ql^+},m_{ql^-}\}
\end{align}

As squarks are pair-produced at the LHC, they decay via the above or
even simpler patterns leading to final states that may be able to mimic those of the leptoquarkino signals: two 
hard partonic jets, two or more opposite sign, same
flavour (OSSF) leptons and most importantly large amounts of missing
transverse energy. However, it is crucial to note, that due to the
general assumption of Minimal Flavour Violation in the SUSY lepton sector,
flavour is a conserved quantum number and runs like a line through the
cascade chain. Hence, OSSF leptons originating from squark decays
inside a single decay chain are (up to a simple combinatorical factor)
equally likely than opposite sign different flavour (OSDF) lepton
contributions for standard MSSM-type signals, since the branching
ratios into first and second generation fermion/sfermion combinations
are of the same size in most mSUGRA scenarios. These OSDF signals are
coming in the standard MSSM paradigm from different flavour decays of
pair-produced SUSY particles. They are, however, not present in the 
case of leptoquarkino decays, when we consider flavour-diagonal Yukawa
couplings and standard FCNC constraints, which allows to suppress
large amounts of standard SUSY 'backgrounds'. 

As a selection criterion, we accept events only if they have exactly
two hard jets with a minimum $p_T$ of 50 \GeV. This requirement is
intended to suppress one additional collinear initial state 
radiation jet, for allowed jet multiplicities of three. If we
also choose to take jet multiplicities of four into account,
we have to be more careful: to avert the contamination by
e.g. a gluino pair production signal, we need the third hardest jet to
have a maximum $p_T$ of 50 \GeV, while retaining the cuts above for
the two hardest jets. In the cases investigated here this was
sufficient to separate leptoquarkino signals from gluino pair
production, however, there is still the possibility from mixed
gluino-squark production which is known to be able to be accompanied
by possibly harder QCD radiation jets~\cite{Plehn:2005cq}.
Furthermore, it could also appear that jets from decays could be
rather soft (e.g. in SPS1a where gluinos and squarks are quite
degenerate), leading to a confusion with QCD radiation jets. For
leptoquarkino signals this seems not to be overly likely (because the
leptoquarkino renormalization group equations are not so tightly
linked as those of squarks and gluinos), but is nevertheless
possible. For the further analysis presented here, we just take the
rather simplified cut above and assume that the jet backgrounds are
under control. A complete background study is beyond the intent of the
present study and is not performed here.  

\begin{figure}
\centerline{
 \includegraphics[width=0.45\textwidth]{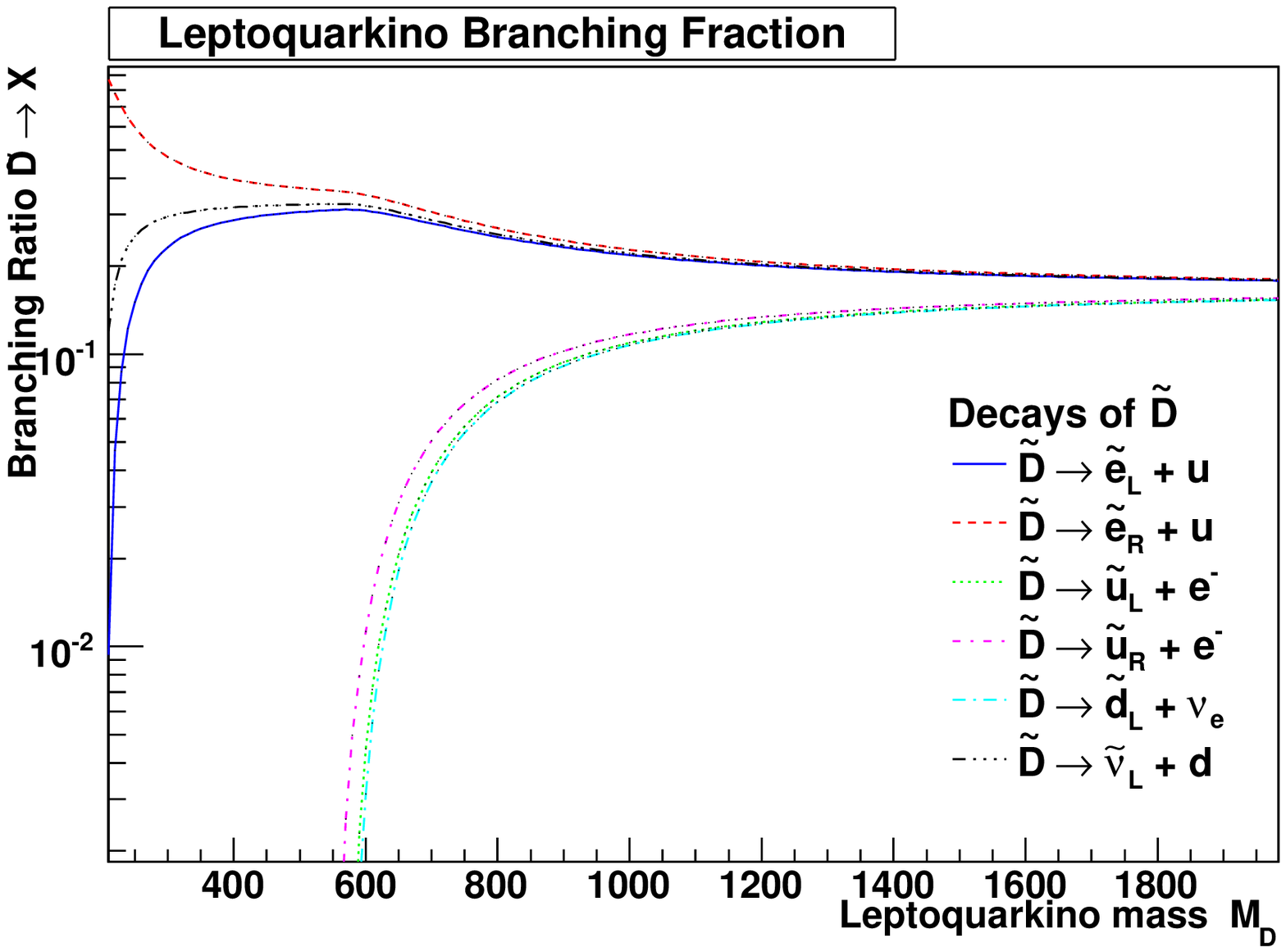}
 \includegraphics[width=0.45\textwidth]{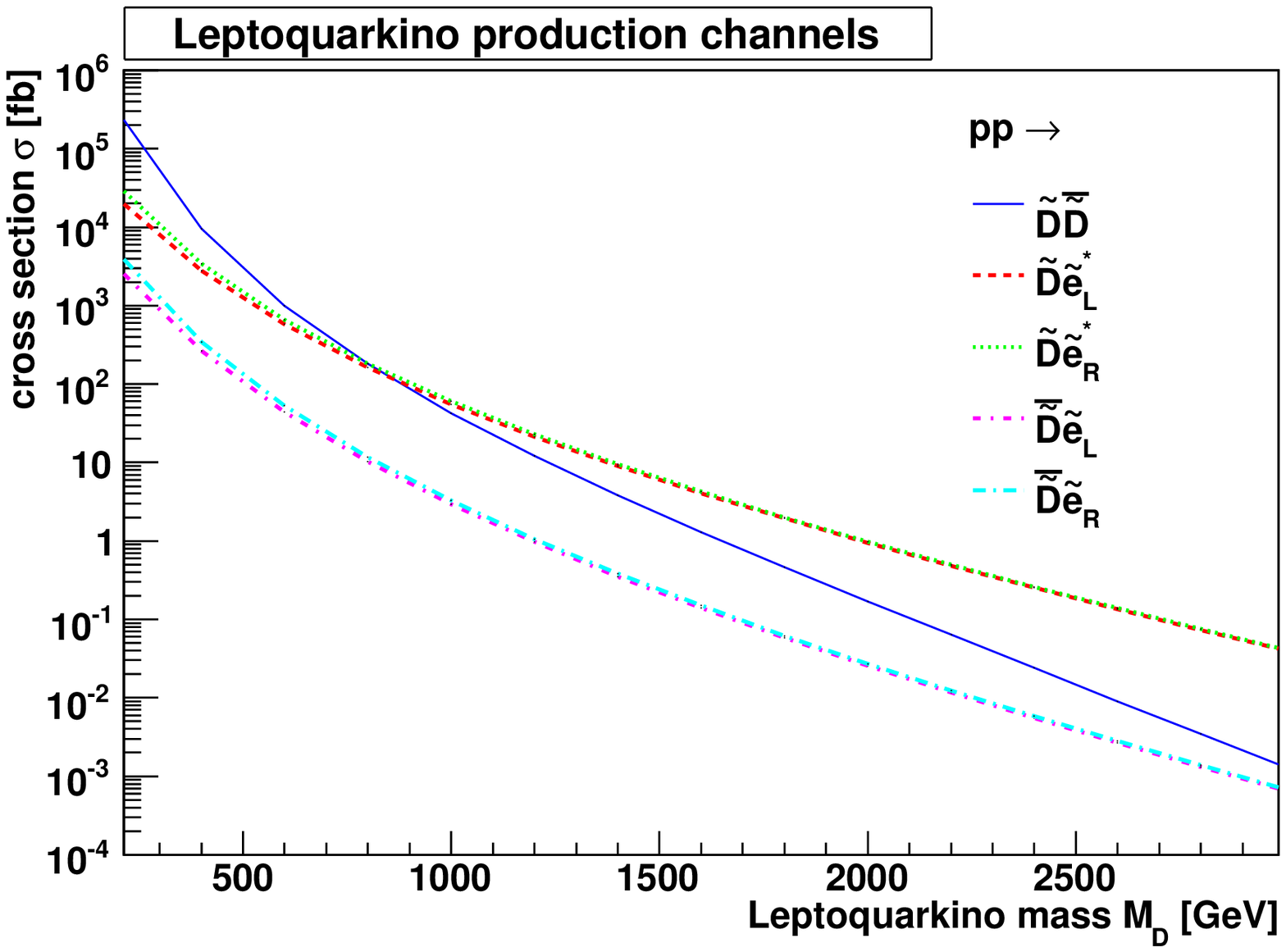}}
 \caption{    \label{cs:br}Branching ratio for the leptoquarkino decay into fermion
 and scalar (left), leading order cross sections for single and pair
 production at 14 \TeV\ (right).} 
\end{figure}

Leptoquarkinos, if existent, are abundantly produced at the LHC, since they are
massive colored \hyphenation{ iso-sing-let } isosinglet fermions
\cite{Hewett:1988xc}. The leading order cross sections basically 
depend only on the mass for pair production, and the Yukawa coupling for
single production, respectively (see RHS of Fig. \ref{cs:br}). The
Yukawa coupling is without knowledge of the complete GUT model
arbitrary, but was taken here to be of the size of the electromagnetic 
coupling ($y = 0.312$). The LHS of Fig. \ref{cs:br} shows the
branching fractions of the decaying leptoquarkino (for varying masses)
into a fermion/sfermion pair. As the decay into squarks
and leptons is kinematically forbidden for low leptoquarkino masses
(and still heavily phase-space suppressed for increasing leptoquarkino masses), 
the sleptons dominate in that case as intermediate states in
cascades. Though this depends on the parameter space of the models, we
follow this assumption from here on. Consequently, a typical
leptoquarkino decay may be given by  
\begin{align}
\tilde{D} \rightarrow q\tilde{l}^-_{R/L} \rightarrow ql^-\tilde{\chi}^0_1 \; ,
\label{eq:lqinocasc1}
\end{align}
whereas a second-to-lightest neutralino in the decay chain starts to
become important for heavier masses, leading to different intermediate states,
e.g.: 

\begin{align}
\tilde{D} \rightarrow \tilde{q}^-_{R/L}l^- \rightarrow
ql^-\tilde{\chi}^0_2 \rightarrow 
ql^-l^\pm l^\mp\tilde{\chi}^0_1 \, .
\label{eq:lqinocasc2}
\end{align}
\begin{figure}[h!]
  \centering
  \includegraphics[width=0.4\textwidth]{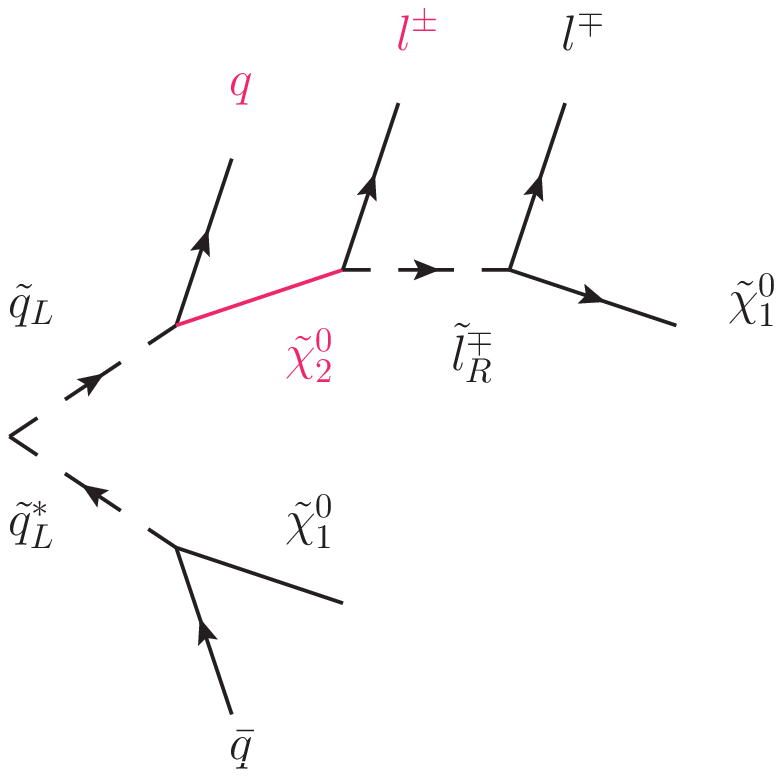}
  \includegraphics[width=0.3\textwidth]{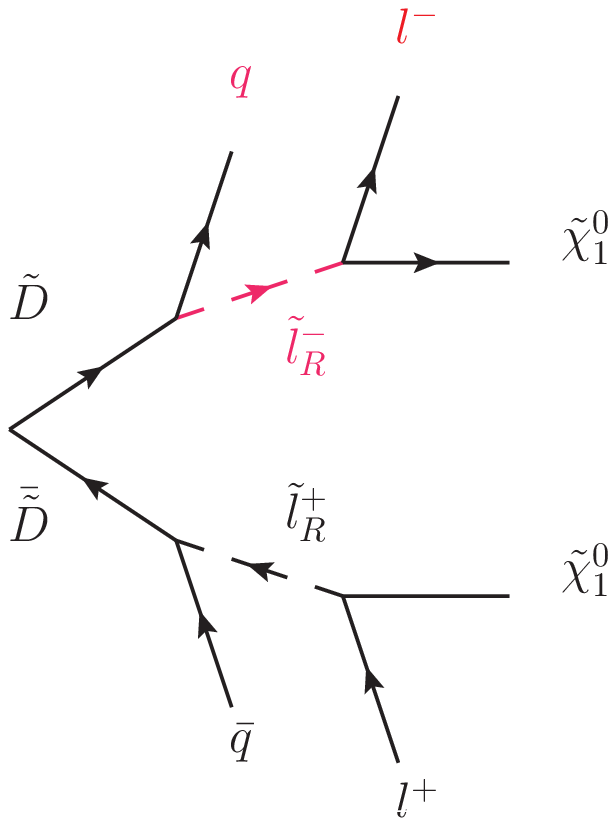} 
  \caption{Examples for decay cascades under investigation: squark
  (left) and leptoquarkino (right) pair production.}
  \label{fig:cascade}
\end{figure}

Leptoquarkinos produced in pairs thus show the same exclusive final
states as squarks, namely two hard partonic jets, two or more leptons
and large missing transverse energy in the detector, accounting for
the undetected neutralinos. The influence of the second type of
cascade (\ref{eq:lqinocasc2}) however is of less importance, since the
new Majorana decay gives rise to additional OSDF lepton contributions,
which is absent in the first type of leptoquarkino
chain. Additionally, heavier intermediate states are strongly
phase-space suppressed and only start to become more important for
increasing exotic masses. Consequently, we focus our analysis on the
first cascade (\ref{eq:lqinocasc1}) with the case of only two OSSF
leptons being present, as it proves to be the most common exotic decay
pattern for the regions in parameter space with relatively light
leptoquarks.

\begin{figure}[h!]
    \centering
      \includegraphics[width=0.45\textwidth]{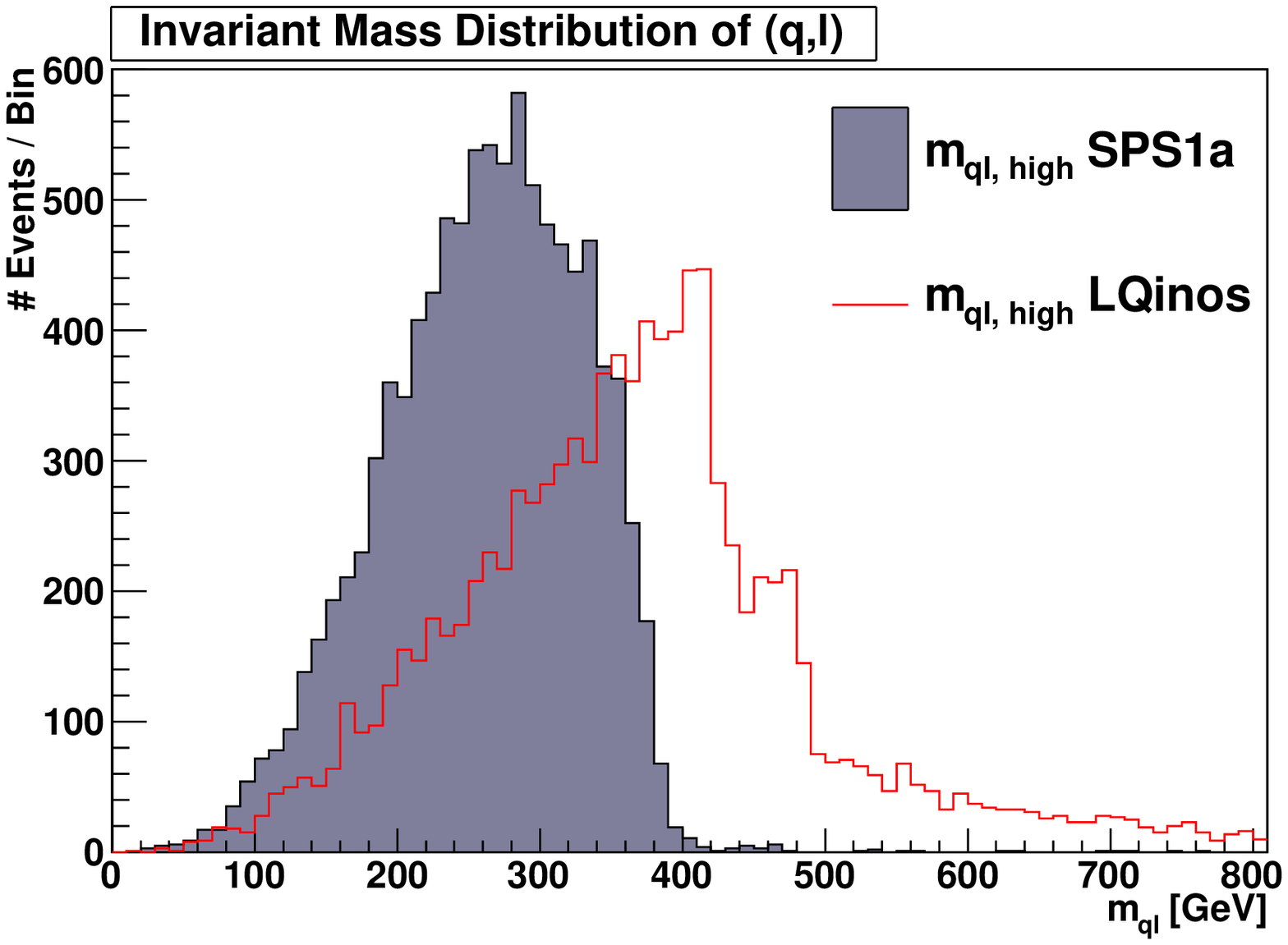} 
      \includegraphics[width=0.45\textwidth]{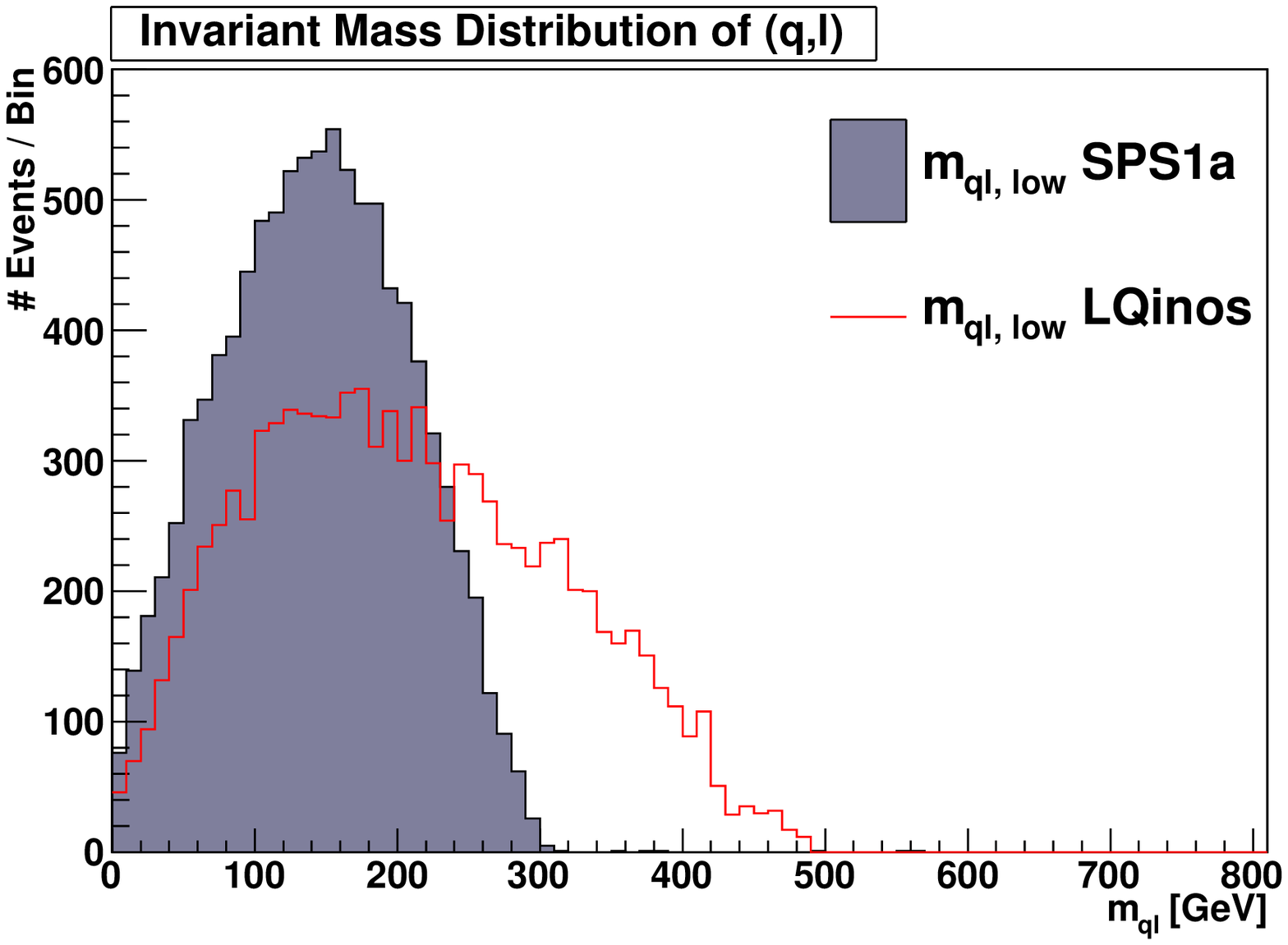} 
  \caption{Anatomy of leptoquarkino mass edges for $m_{ql,high}$ and
  $m_{ql,low}$ with $m_D = 600 \enspace \GeV$} 
  \label{mlq:highlow}
\end{figure}
At this early partonic stage, standard analysis methods for
lepton-quark mass edges applied to events with dedicated leptoquarkino
cascades show strong discrepancies to well-known results from standard
SUSY signals. The difference emerges due to the intermediate on-shell
scalar (squark or slepton) between the quark and lepton 
compared to a Majorana fermion as e.g. the neutralino in the MSSM:
there are no possible spin correlations between lepton and quark, as
they are connected through a scalar propagator. As a result, their
invariant mass spectrum is equivalent to the dilepton spectrum in
standard MSSM models (stemming from a scalar slepton propagator), in
that it linearly rises from zero to its maximum at the endpoint, where
it instantly falls down to zero. This case is not necessarily a unique
feature of SUSY as there are other models with a scalar propagator,
which may distort the actual shape of the relevant invariant mass
variables (e.q. in UED, for a general review cf
\cite{Athanasiou:2006ef}). However we take the leptoquarkino setup as
an application prototype and investigate how such deviations from
possible intermediate scalars can arise in the context of a SUSY decay
chain. A complete knowledge of the particular extended SUSY model
containing leptoquarkinos is not needed for the present study, but
will be discussed in a future publication, which in addition will
discuss a set of discrete self-consistent parameter points
\cite{Braam:2011}. The generic properties of leptoquarkinos, however,
could be inferred rather independent of the underlying model setup:
they are scalar particles which decay into final states with lepton
and baryon number, but because of the assumption of conserved R
parity do not show a peak in the corresponding invariant mass distribution. 
The smoking-gun signatures are the specific quark-lepton mass edges to be
described below in detail, which deviate from standard edges in SUSY
squark and gluino searches. (Such decays could indeed arise from
squarks in SUSY models with $R$ parity violation, however, due to
constraints from proton lifetime these would be overwhelmingly washed
out by the standard SUSY decays). The flavor structure of
leptoquarkinos might be quite interesting, especially when considering
third-generation leptoquarkinos involving (s)tops, (s)bottoms
(experimentally problematic because of the neutrinos in the final
state) and (s)taus, but is rather model-dependent and thus we postpone 
a more comprehensive analysis to~\cite{Braam:2011}.

For the comparison of ordinary squark- with leptoquarkino cascades we
are exemplarily using the parameter point \textit{SPS1a}~\cite{sps}
for the MSSM 
as well as a model containing leptoquarks\footnote{The scalars are
  considered heavier than fermions (masses well above 1 TeV), since
  this is usually the case and their presence would most likely alter 
  the shape in that a resonant peak structure would dominate the
  spectrum.} 
and -inos with varying masses augmented by squarks and sleptons with
the same masses as the \textit{SPS1a} data point ($m_{\tilde{u}_L} =
567$ \GeV, $m_{\tilde{u}_R} = 547 $  \GeV, $m_{\tilde{l}_L} = 204 $
\GeV, $m_{\tilde{l}_R} = 145 $ \GeV, $m_{\tilde{\chi}^0_1} = 97 $
\GeV, $m_{\tilde{\chi}^0_2} = 181 $ \GeV). To reduce the dependency of
the effect of edge distortion on the SUSY scenario, we also looked at
three other heavy Snowmass Points and Slopes (SPS) spectra including
one gauge-mediated symmetry breaking (GMSB) case (\textit{SPS1b, 
SPS3, SPS7}~\cite{sps}), the results of which are summarized in the
appendix. In general, we find that the following analysis does not
depend on the SUSY breaking scenario or the detailed specifics of the
chosen parameter point. 

\begin{figure}
    \centering
      \includegraphics[width=0.45\textwidth]{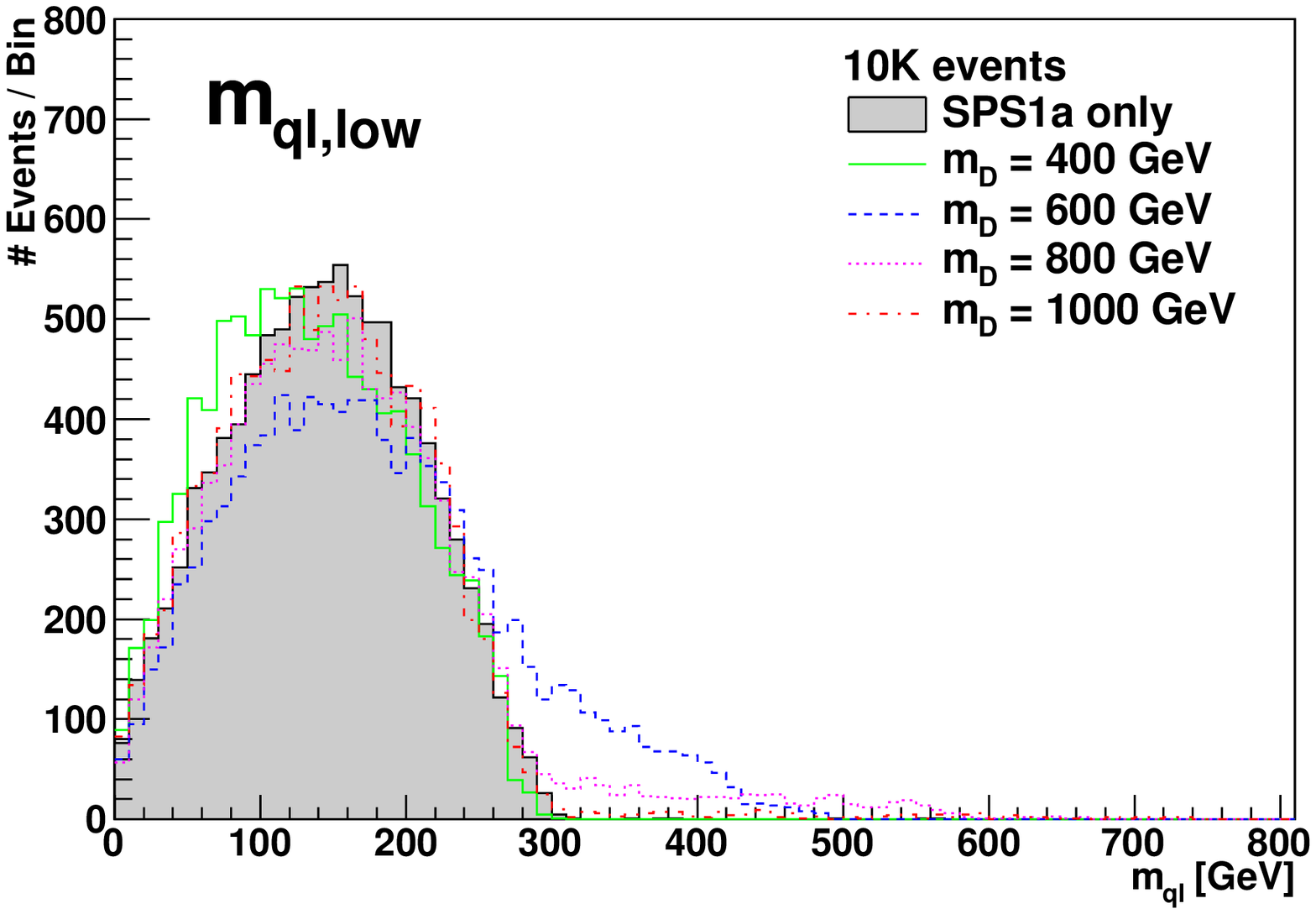} 
      \includegraphics[width=0.45\textwidth]{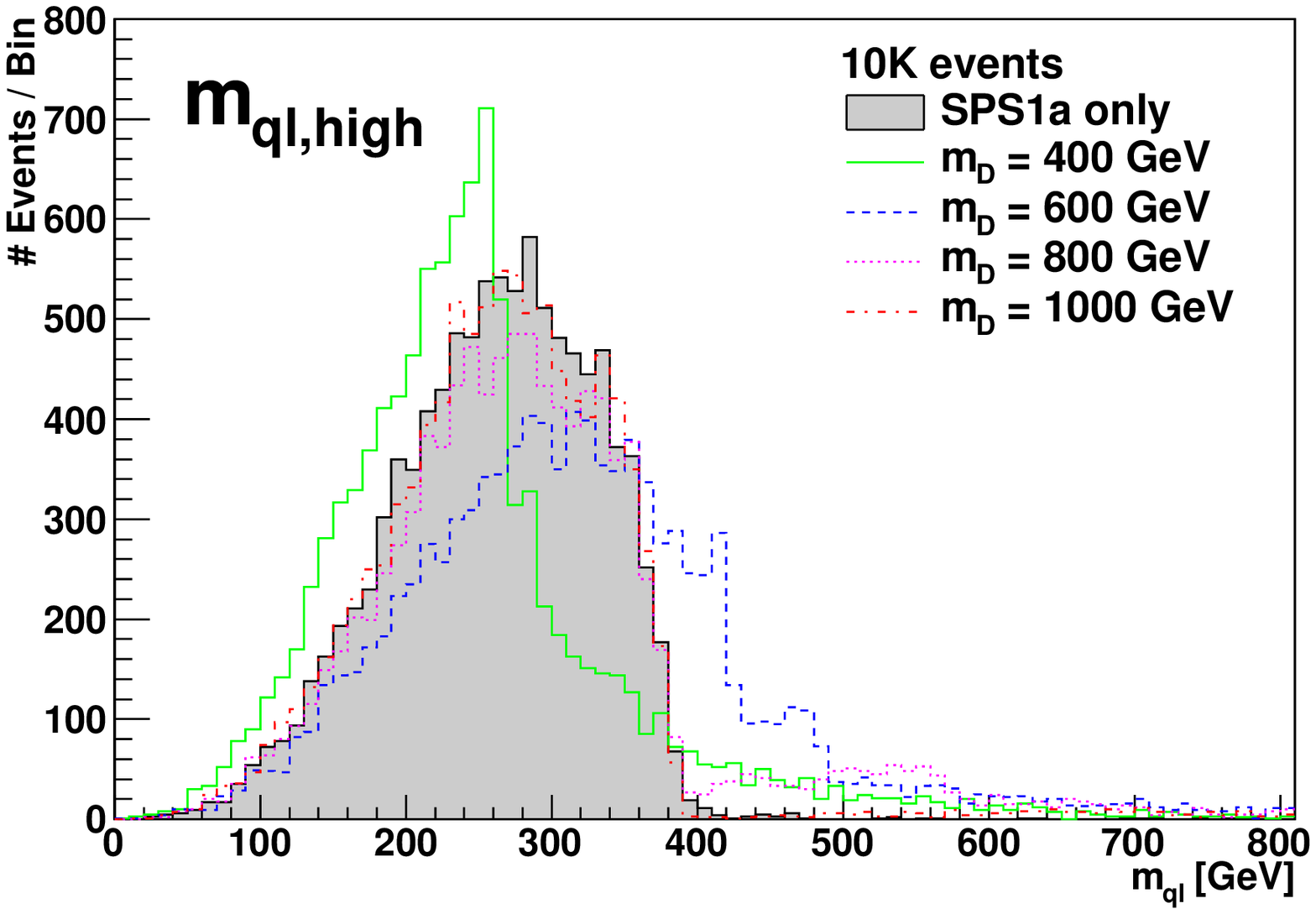} \\
      \includegraphics[width=0.45\textwidth]{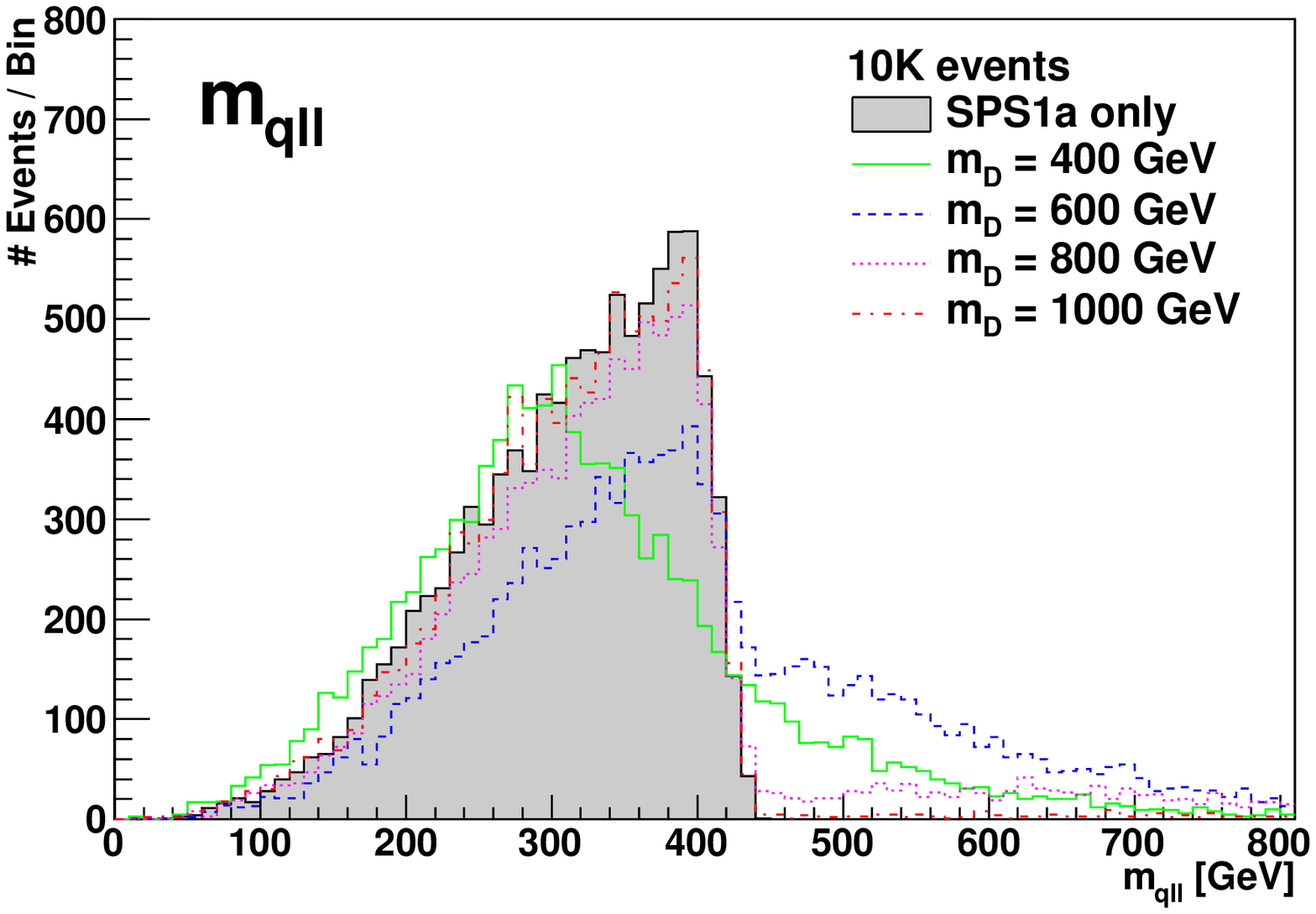} 
      \includegraphics[width=0.45\textwidth]{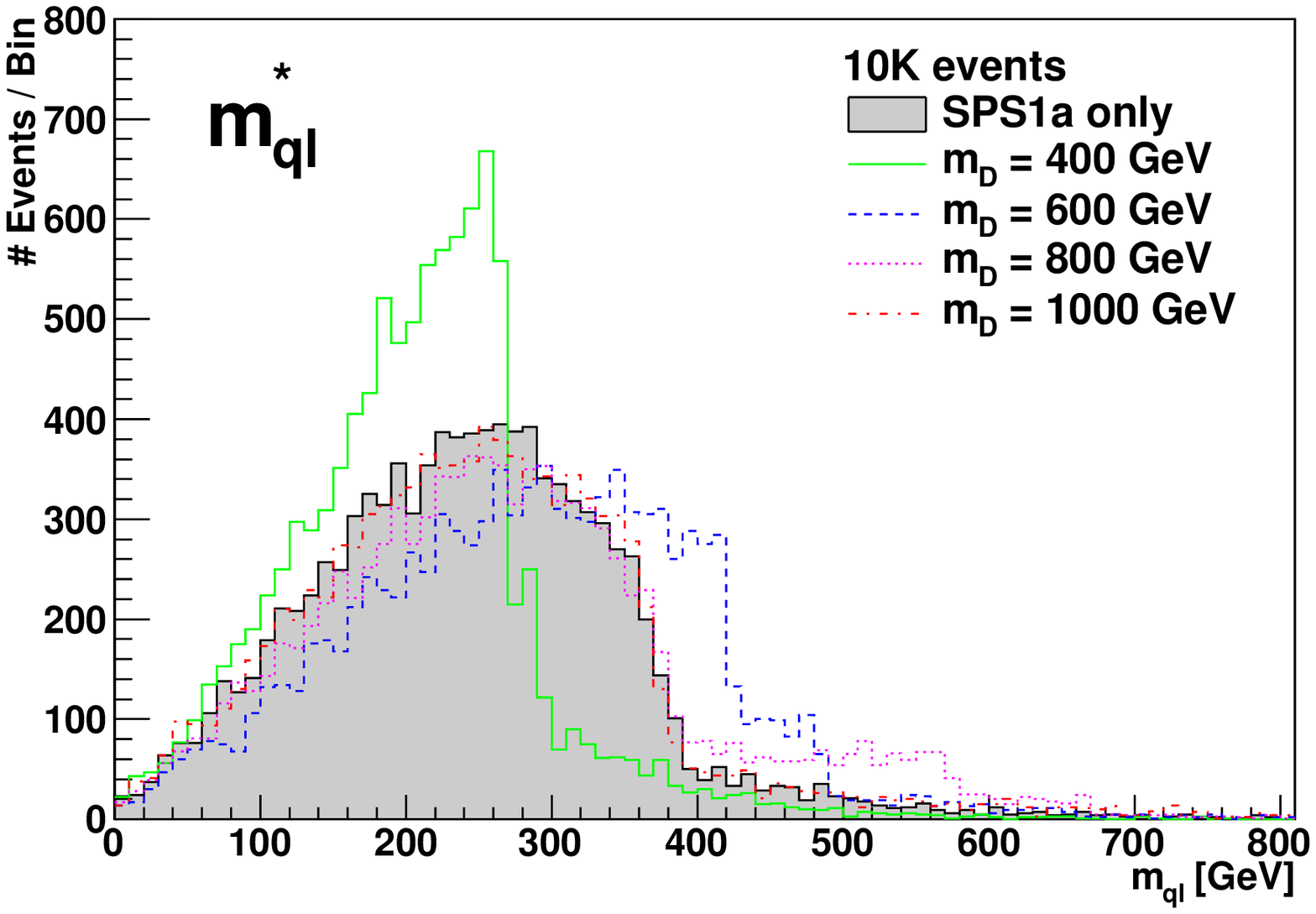} 
  \caption{Leptoquarkino mass scans between 400 and 1000 GeV show
  deviations from standard SUSY observables in an underlying SPS1a
  spectrum. The upper left and right figures disclose the quark-lepton
  invariant mass spectra of $m_{ql,low}$ and $m_{ql,high}$,
  respectively. The lower parts show the complete quark-dilepton
  spectrum $m_{qll}$ on the left and the combinatorics-free $m_{ql}^*$
  on the right, which is defined in
  Eqn. (\ref{eq:mlqstern}). Continuous, dashed, dotted and dash-dotted
  lines correspond to leptoquarkino masses of 400, 600, 800 and 1000 GeV,
  respectively.}
  \label{fig:scan}
\end{figure}

\begin{figure}
    \centering
      \includegraphics[width=0.45\textwidth]{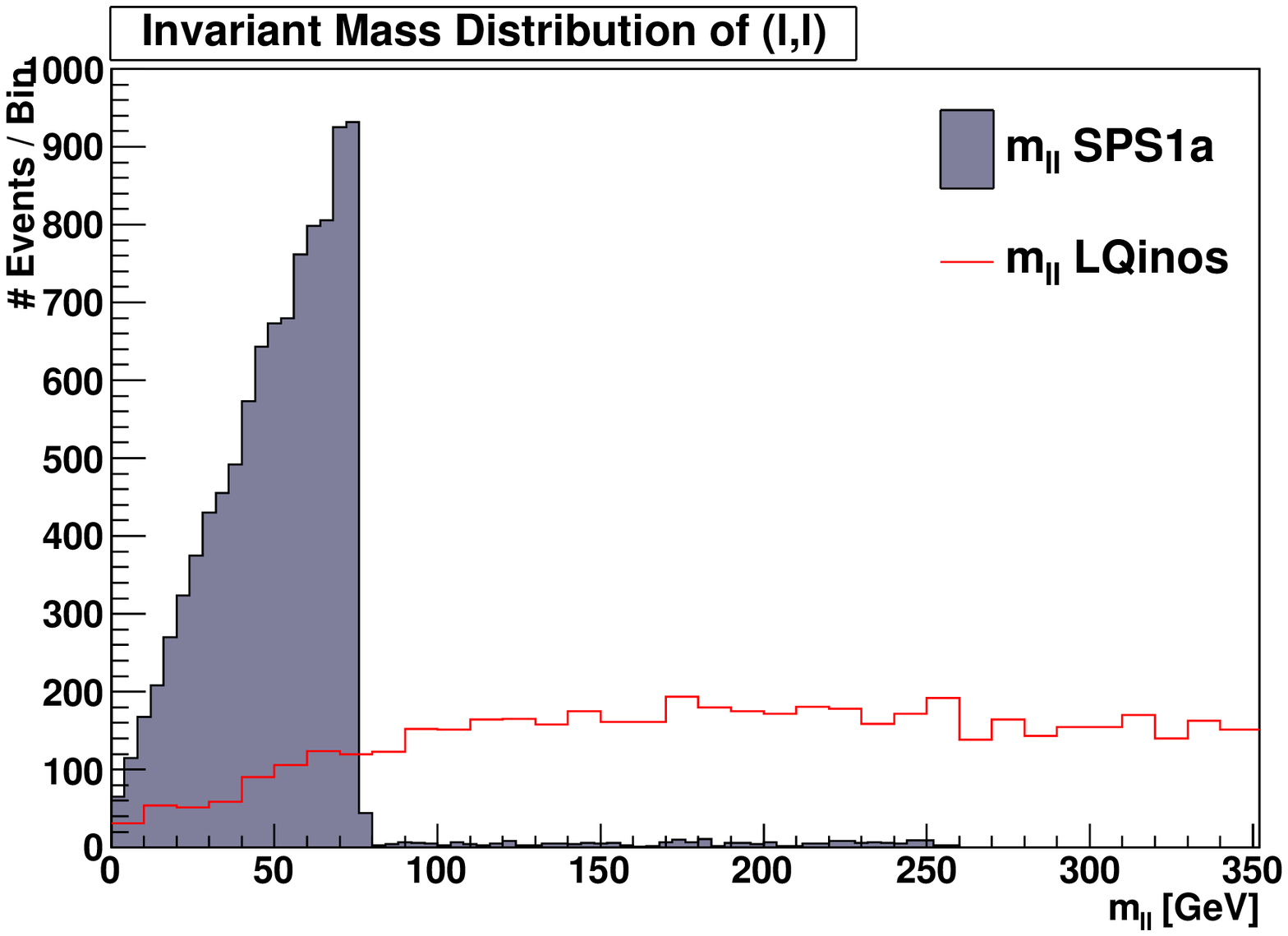} 
      \includegraphics[width=0.45\textwidth]{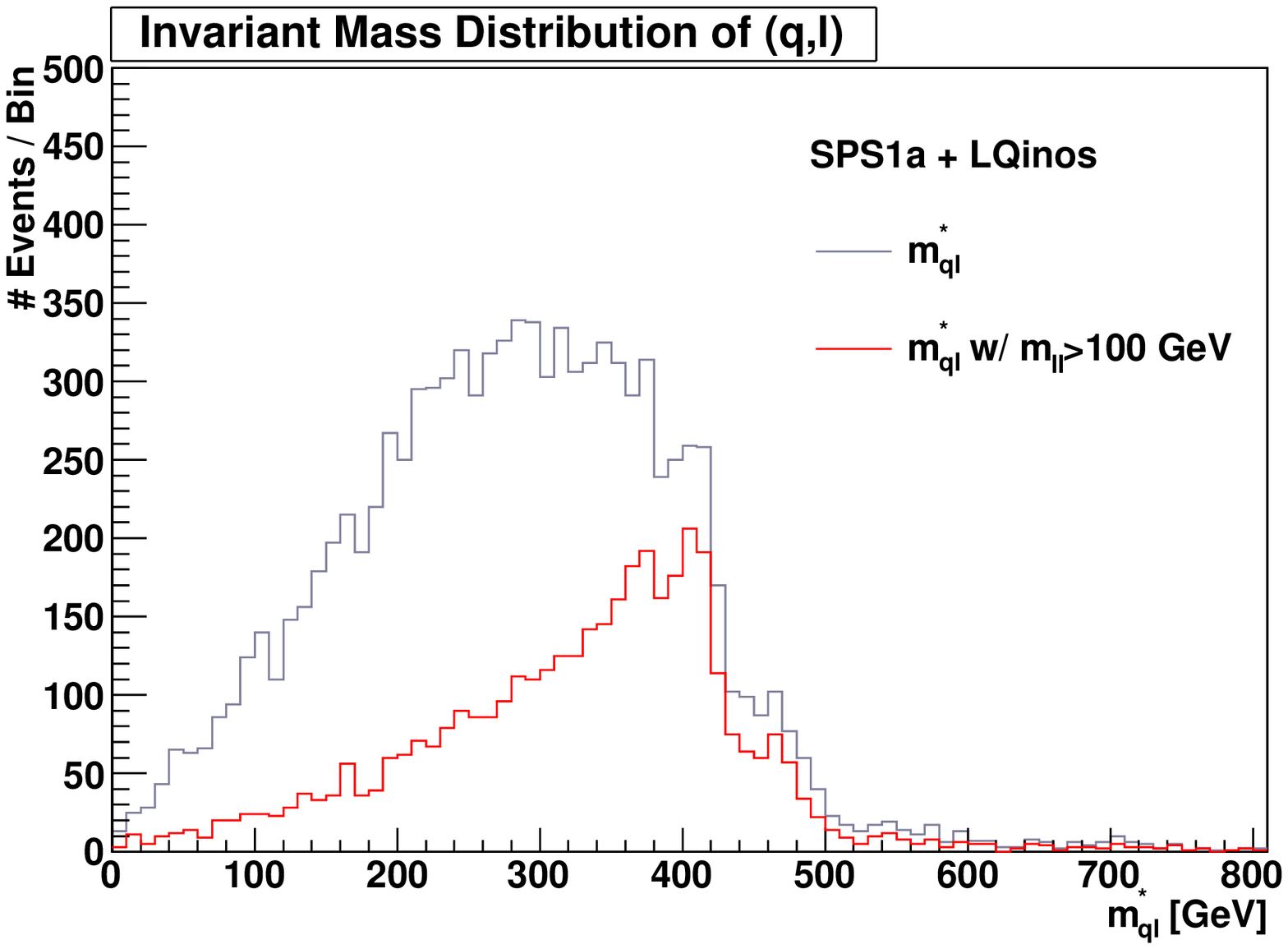} 
  \caption{The LHS shows the dileptonic spectrum with strong
  discrepancies between pure SPS1a squark- and leptoquarkino signals
  in that the correlation of both leptons from one specific decay
  chain is obvious in the first case. This can be used to severely
  suppress the standard SUSY signals by cutting on $m_{ll}$ shortly
  above the kinematical edge. The RHS presents the impact of this cut
  on the variable defined in Eqn. (\ref{eq:mlqstern}) ($m_D =
  600\enspace \GeV$).}
  \label{fig:invmassll}
\end{figure}

The kinematical endpoints under consideration are given by the masses
of the intermediate and mother particles:  
\begin{align}
 m^{max}_{ql} = \enspace & \left[ \frac{(m_{\tilde{e}_{R(L)}}^2 -
 m_{\tilde{\chi}^0_{1}}^2)(m_{D}^2 -
 m_{\tilde{e}_{R(L)}}^2)}{m^2_{\tilde{e}_{R(L)}}}
 \right]^{\frac{1}{2}}\nonumber \\  
 = \enspace &  433 \enspace (496) \enspace \GeV
  \label{eq:sel_edges2}
\end{align} 
for a leptoquarkino mass of 600 GeV, where the value in parentheses is
given for an intermediate left-handed slepton, which is slightly
phase-space suppressed in this particular \textit{SPS1a} scenario. The overall signal consists of the sum of both
contributions leading to the shape visible in Figures
\ref{mlq:highlow}, \ref{fig:scan} and \ref{fig:invmassll}. 
For each model, a data set of 10K unweighted events was generated using a hard-coded
implementation of these $E_6$-inspired SUSY models into the event
generator {\sc Whizard}~\cite{whizard}, which is particularly well
suited for LHC beyond the SM studies~\cite{whizard_bsm}. While a
complete validation of the model implementation using the {\sc
  Whizard} interface to the {\sc FeynRules} package~\cite{feynrules}
is under way, the part of the implementation relevant for this paper
has been extensively tested. \\

Returning to the cascade, there still remains the problem of
observability: experimentally there is no possibility to select the
correct partonic jet and corresponding lepton, which are then to be
combined to the invariant mass spectrum. While in MSSM models this
would come about due to the presence of the Majorana fermion decay
into two OSSF leptons, in the leptoquarkino case two OSSF leptons are
to be collected from different cascades, one originating from the
leptoquarkino and the other from its antiparticle, respectively. The
observables $m_{ql,low}$ and $m_{ql,high}$, shown in
Fig. \ref{mlq:highlow}, thus display the tremendous discrepancy, especially
the latter one with its sharply falling edge shape, intrinsic to the
nature of the scalar intermediate sparticle. The long tail, inherent
in the signal variant of $m_{ql,high}$, is another unusual feature and
a direct consquence of the combination of final state particles from
two different cascade sides.

This issue of combinatorics however may be addressed by combining the
softest jet and the hardest lepton to form an invariant mass spectrum
in a single event. This has proven to be useful \cite{Kang:2007ib} in
terms of resembling the actual shape and thus the most accurate
position of the theoretical edge:
\begin{align}  
  m_{ql}^* = m(\min_E \{j_1,j_2\},\max_E \{l^+,l^-\})
 \label{eq:mlqstern}
\end{align}

\begin{figure}
    \centering
      \includegraphics[width=0.45\textwidth]{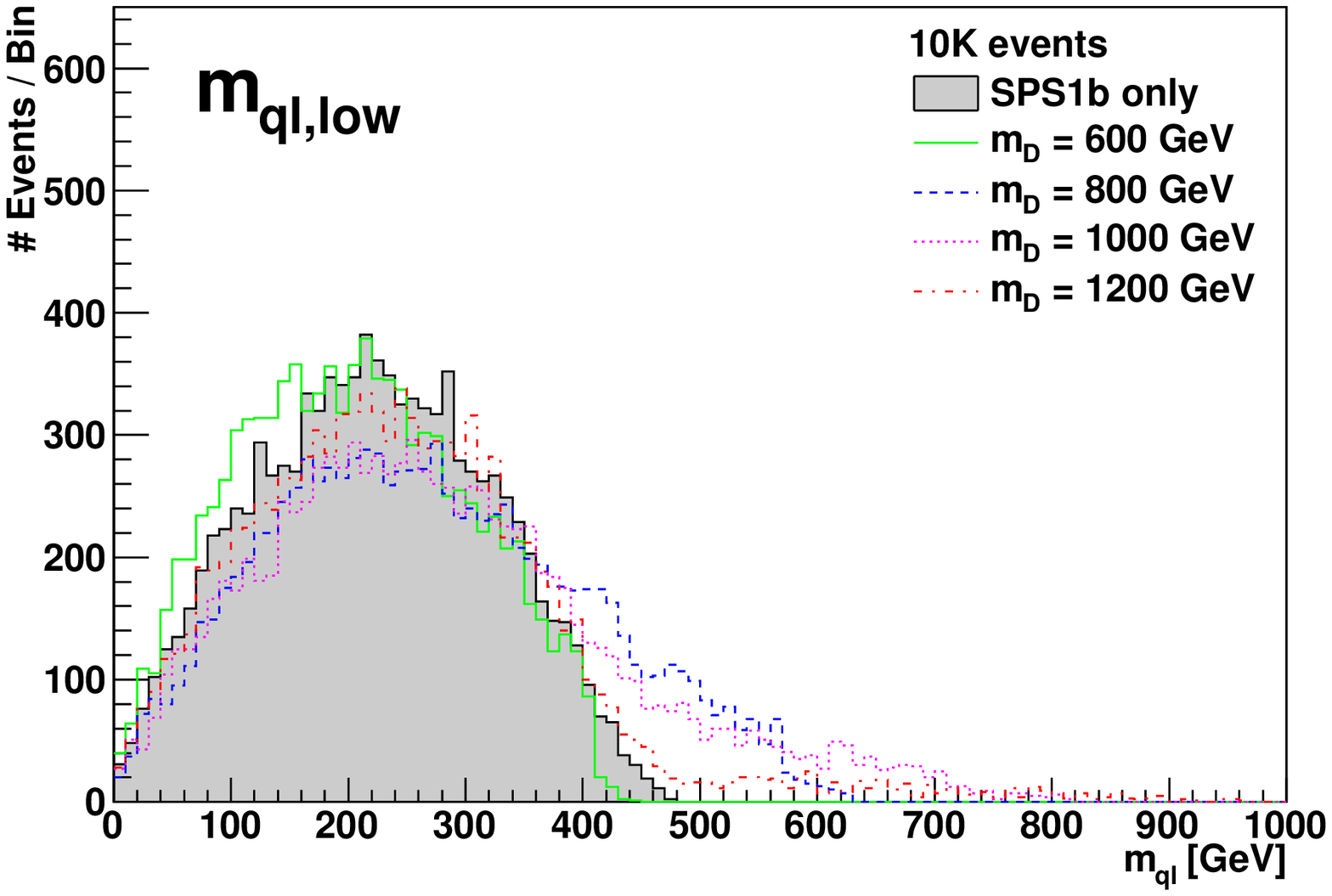} 
      \includegraphics[width=0.45\textwidth]{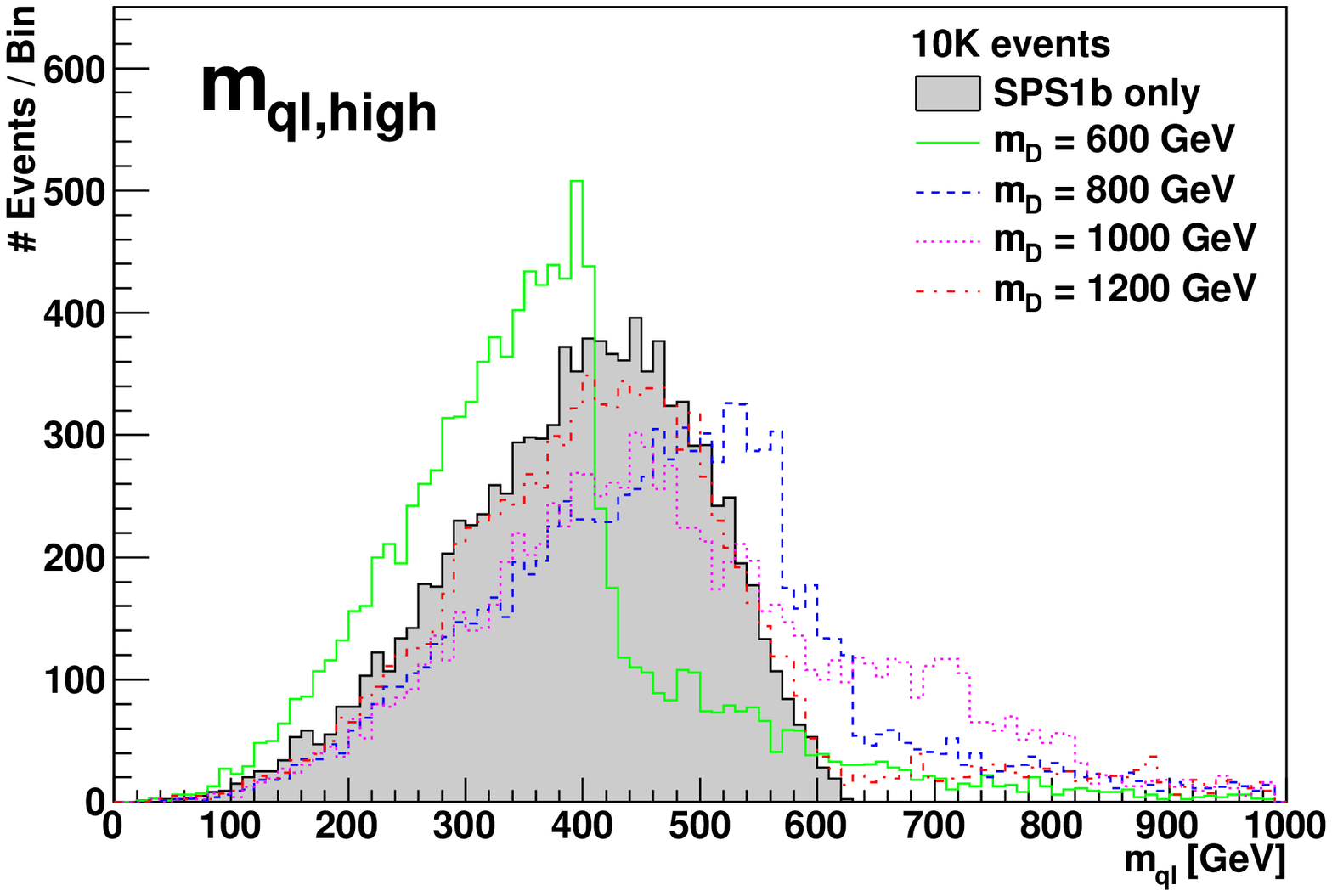} \\
      
      \vspace{-4mm}

      \includegraphics[width=0.45\textwidth]{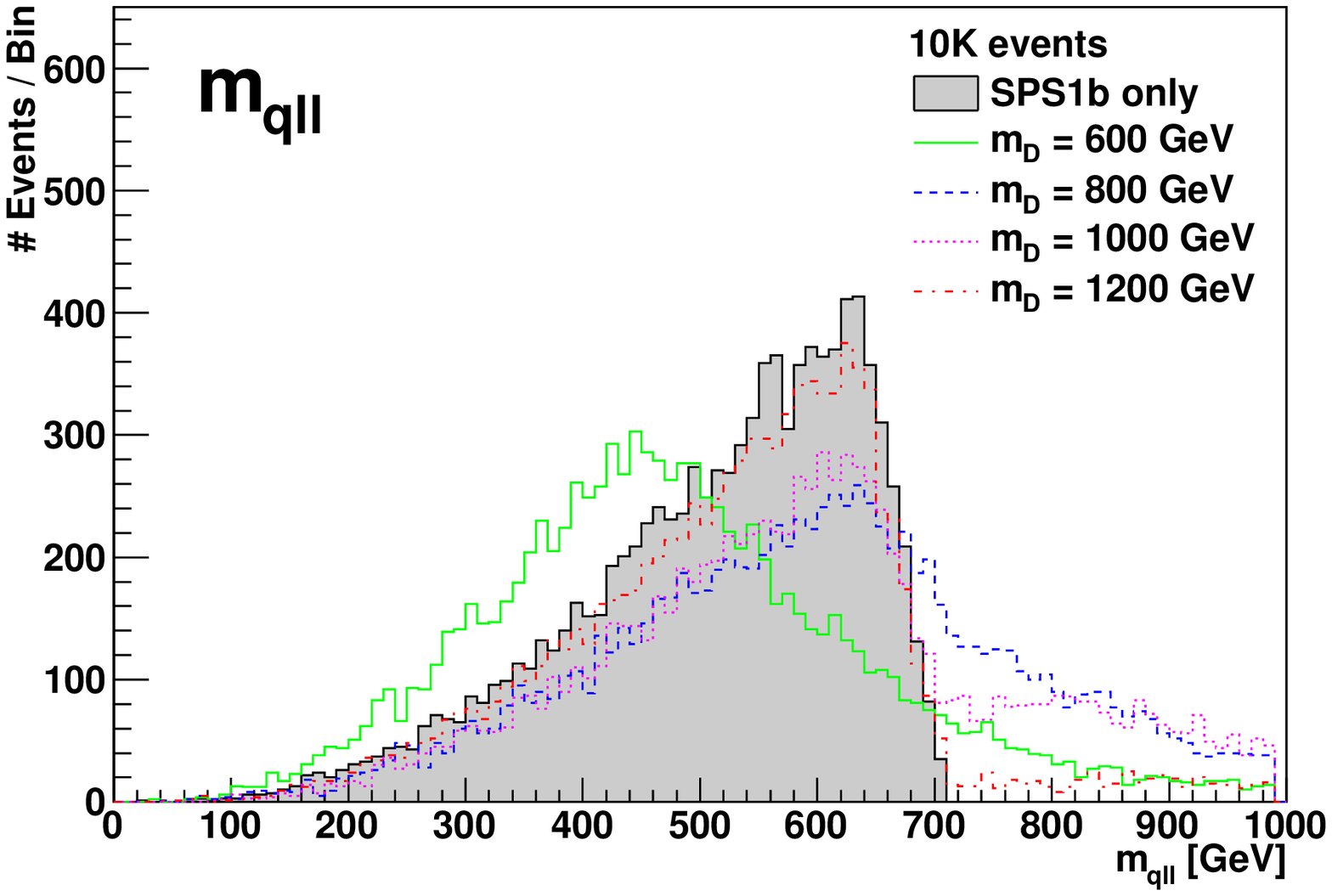} 
      \includegraphics[width=0.45\textwidth]{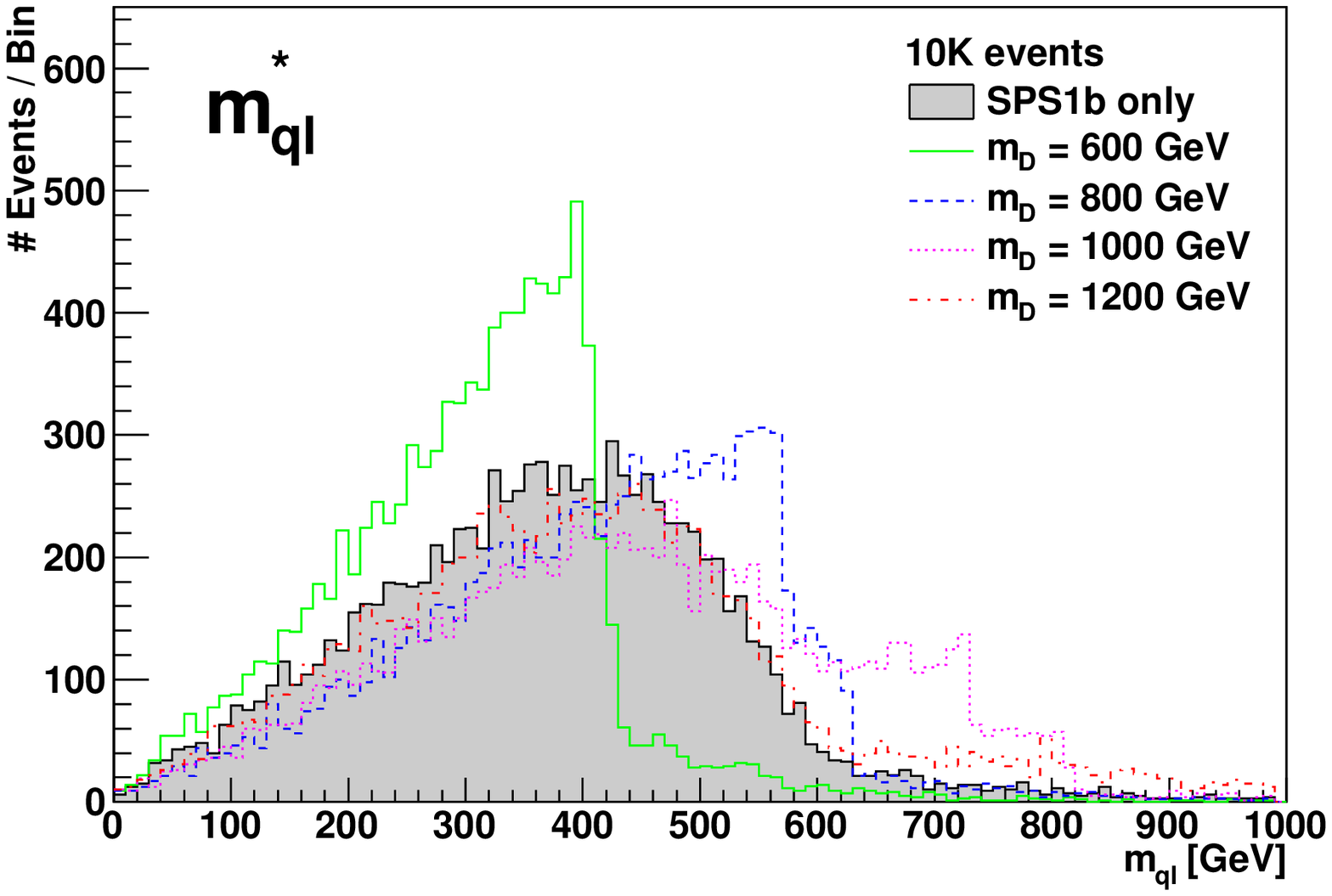}
      \caption{The same variables as in \ref{fig:scan}, instead
        dressed with an underlying SPS1b. Continuous, dashed, dotted
        and dash-dotted lines correspond to leptoquarkino masses of
        600, 800, 1000 and 1200 GeV, respectively.} 
  \label{fig:scan2}
\end{figure}

This observable is analysed in Fig.~\ref{fig:scan} for \textit{SPS1a}
(the other SPS scenarios are found in
Figs.~\ref{fig:scan2}-\ref{fig:scan4}) together with  
$m_{ql,high}$, $m_{ql,low}$ and $m_{qll}$ for four different
leptoquarkino masses ranging from 400 \GeV $\enspace$to 1000 GeV
embedded into an underlying SPS1a spectrum. Since the hadronic leading
order cross-section for pair produced leptoquarkinos is of the order
of $10^{4}$ fb for masses as low as 400 GeV (see Fig~\ref{cs:br}), it
dominates the shape of the observables in Fig.~\ref{fig:scan}. 
The specific shapes of the observables directly hints towards new,
possibly exotic physics beyond a standard SUSY paradigm. However, this 
behaviour changes for heavier leptoquarkinos: the higher the mass and
thus the lower the cross-section of the involved leptoquarkino, the
weaker the effect on the observables. While for masses of 600 GeV the
shape of the distribution is still predominantly given by the
leptoquarkino contribution, differences to squark signals are merely
visible for 800 GeV and are practically invisible for 1000 GeV (or
will be so after detector effects). These characteristics do not change
drastically for the other SPS spectra but are instead shifted to
higher values of the leptoquarkino mass\footnote{Mass scans are
  adjusted to take place from 600 to 1200 \GeV}. The feature of edge
distortion persists in all of the different SUSY scenarios and is thus
mainly independent of the latter, as long as the leptoquarkino mass
is not much heavier than the rest of the spectrum.

\begin{table}
\centering
  \begin{tabular}{|r|c|c|c|}
    \hline 
    $m_D$ & \scriptsize{\# N(LQino) \& N(SUSY)} & \scriptsize{\#
    N$_{cut}$} & \scriptsize{S / $\sqrt{\text{S+B}}$}\\
    \hline
    \hline
    400   & 8763 & 5061  & 54 \\
    600   & 1355 & 540  & 15 \\
    800  & 684  & 102  & 4 \\
    1000 & 594  & 24  & 1 \\
    \hline
  \end{tabular}
  \caption{Significance estimates for $100$ fb$^{-1}$ and the relevant
  (non-) standard SUSY final state of two hard jets, two OSSF leptons
  and $\sla{E}_T$. We only consider standard SUSY events as
  possible backgrounds.}
  \label{tab:s2b}
\end{table}

Given that $m_{ql,i}$ ($i = high,low$) and $m_{qll}$ are
single-sided\footnote{i.e. they are to be applied on one side of the
  cascade only} variables, the inclusion of the leptoquarkinos spoils
the clear endpoint structure due to the presence of (at least) one
lepton from the opposite ('wrong') side. A special remark is in place
here: if kinematical endpoints as the ones discussed above are
distorted by a certain amount of events, which are positioned in a
kinematically inaccessible region (i.e. beyond the edge), it may not
necessarily mean, that the combinatorics of a 'typical' squark
analysis went berserk, but rather indicate an underlying process, that
has, as in our case, exotic fermions involved. A closer look onto a
supplementary observable such as $m_{ql}^*$ from Eqn.
(\ref{eq:mlqstern}), whose construction is free of combinatorical
issues (which, however, could still suffer from an admixture from
different decay sides), may then deliver the additional insight into
the specific process and thus the nature of the model in question.

Furthermore considering one specific SPS scenario, namely
\textit{SPS1a}, together with the 
extra exotic matter content, one is able to discriminate the squark
and leptoquarkino signals by means of the two involved OSSF leptons in
the following way: observation of the dileptonic invariant mass
spectrum may indicate a kinematic endpoint for a vanilla MSSM signal,
which would not be the case for a signature of leptoquarkinos. A rough
significance estimate for $100$ fb$^{-1}$ at 7 TeV thus is possible by
comparing the number of events for the sum of MSSM-type squark and
leptoquarkino cascade to the ones who survive an analysis cut of
$m_{ll}>100$ GeV. The latter is based on the fact that the
kinematical endpoint of the standard SUSY dilepton edge for underlying
\textit{SPS1a} is located below this fixed point in the invariant mass
spectrum.

Significance estimates are given in Tab.~\ref{tab:s2b} while
Fig.~\ref{fig:invmassll} highlights the discrepancies between standard and
non-standard SUSY signals inherent in the dileptonic signal and the
effect of the analysis cut on the variable $m_{lq}^*$ from
Eqn. (\ref{eq:mlqstern}) for a fixed exotic leptoquarkino mass of $m_D =
600$ \GeV. 

\section{Conclusions}

In this letter, we showed in the context of a quite general setup of
GUT-inspired SUSY models containing non-standard SUSY states how the
physiognomy of standard kinematic variables for mass determination of
cascade states can be altered. This happened as the lack of spin
correlations does not distort the shape of those observables. However,
the missing spin correlations do not come from a complete change of the
underlying model paradigm (e.g. assuming Universal Extra Dimensions,
UED) but from a slight variation or extension of the standard SUSY
scenario. The potential for a possible confusion in the model
discrimination is specifically given in the case that the
corresponding scalar partners are too heavy to appear as
resonances at LHC. The significant issue of combinatorics may have a
huge impact on the analysis, if interpreted in the wrong way. Events
in certain regions, which are kinematically forbidden in vanilla MSSM models,
are prone to be dismissed. This comes about due to their false
misidentification based on combinatorics and prevents a possible
disclosure of a real signal from extended SUSY models or the like, in
which the population of the aforementioned regions is
inherent. Consequently this might serve as a prime example to show
that model discrimination at LHC is a crucial, but tedious task, that
needs to be carried out very carefully. Slightest variations of the
underlying model assumptions may distort the shape and appearance of
observables and likely misidentifications may furthermore even prevent
the discovery of new, possibly exotic physics signals. 

Leptoquarkinos in the context of SUSY models is a framework to
consistently embed fermionic leptoquark states in a renormalizable
model. These particles have, depending on their mass, rather high
production rates at the LHC, and their smoking-gun signatures are very
pronounced mass edges in several lepton-jet mass variables compared to
standard squark/gluino mass edges. Other features like mass pattern
and flavor structure are more model-dependent and not generic 
for these states. The discovery of leptoquarkinos could provide a
direct access to the GUT scale structure of supersymmetric field
theories.


\subsubsection*{Acknowledgements}

We thank Alexander Knochel, Felix J\"order and Christoph Horst for
stimulating remarks, Felix Braam for providing us with his code and
Christian Speckner for help with the implementation. 
This work has been supported by the German Research Council (DFG)
under Grant No. RE/2850/1-1 as well as by the Ministery for Research
and Culture (MWK) of the German state Baden-W\"urttemberg, and has
also been partially supported by the DFG Graduiertenkolleg GRK 1102
``Physics at Hadron Colliders''. At the very end of the project, D.W.
acknowledges support from the Scottish Universities Physics Alliance
(SUPA).   

\appendix

\section{Leptoquarkino edges for varying SUSY spectra}

\begin{figure}
      \includegraphics[width=0.45\textwidth]{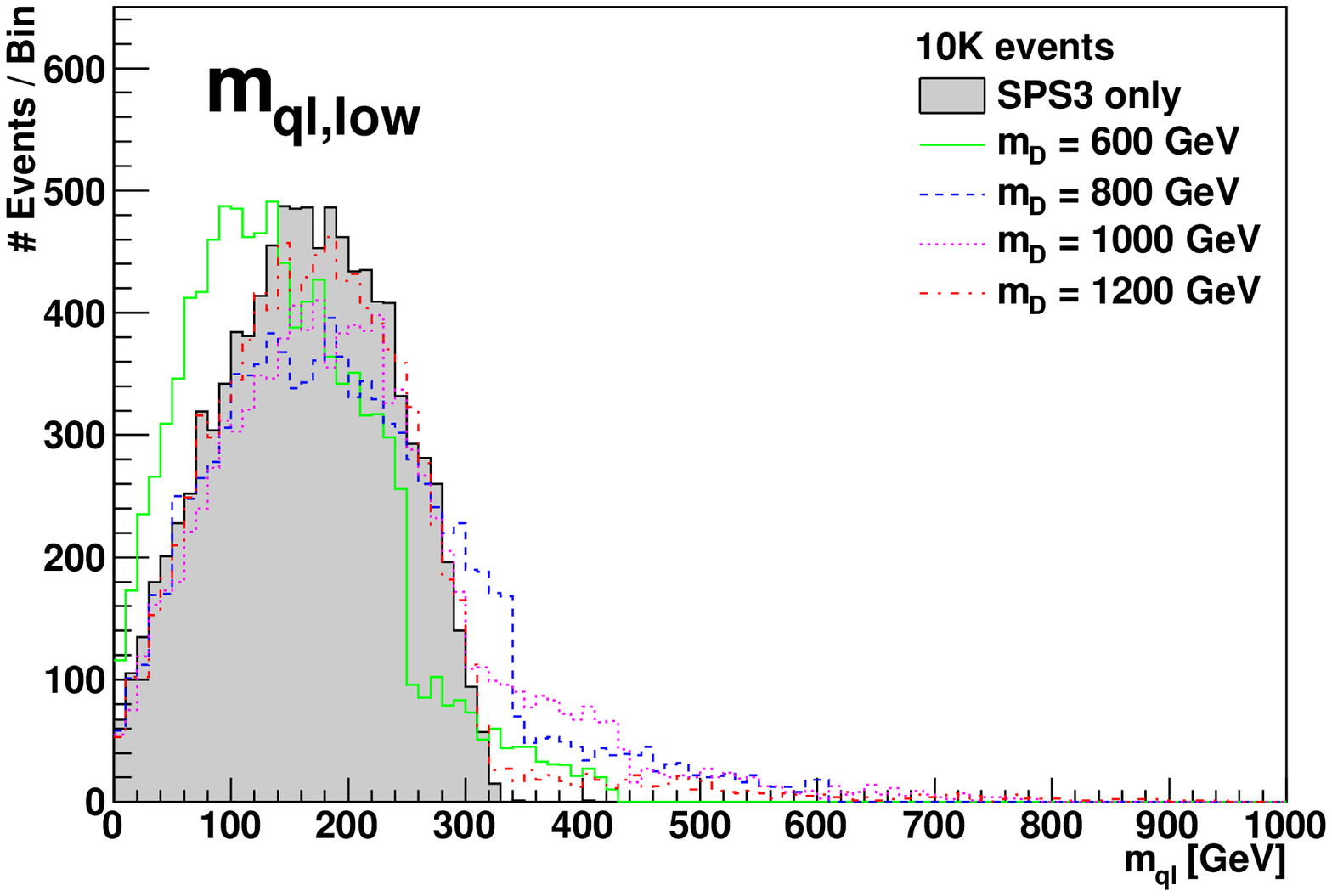} 
      \includegraphics[width=0.45\textwidth]{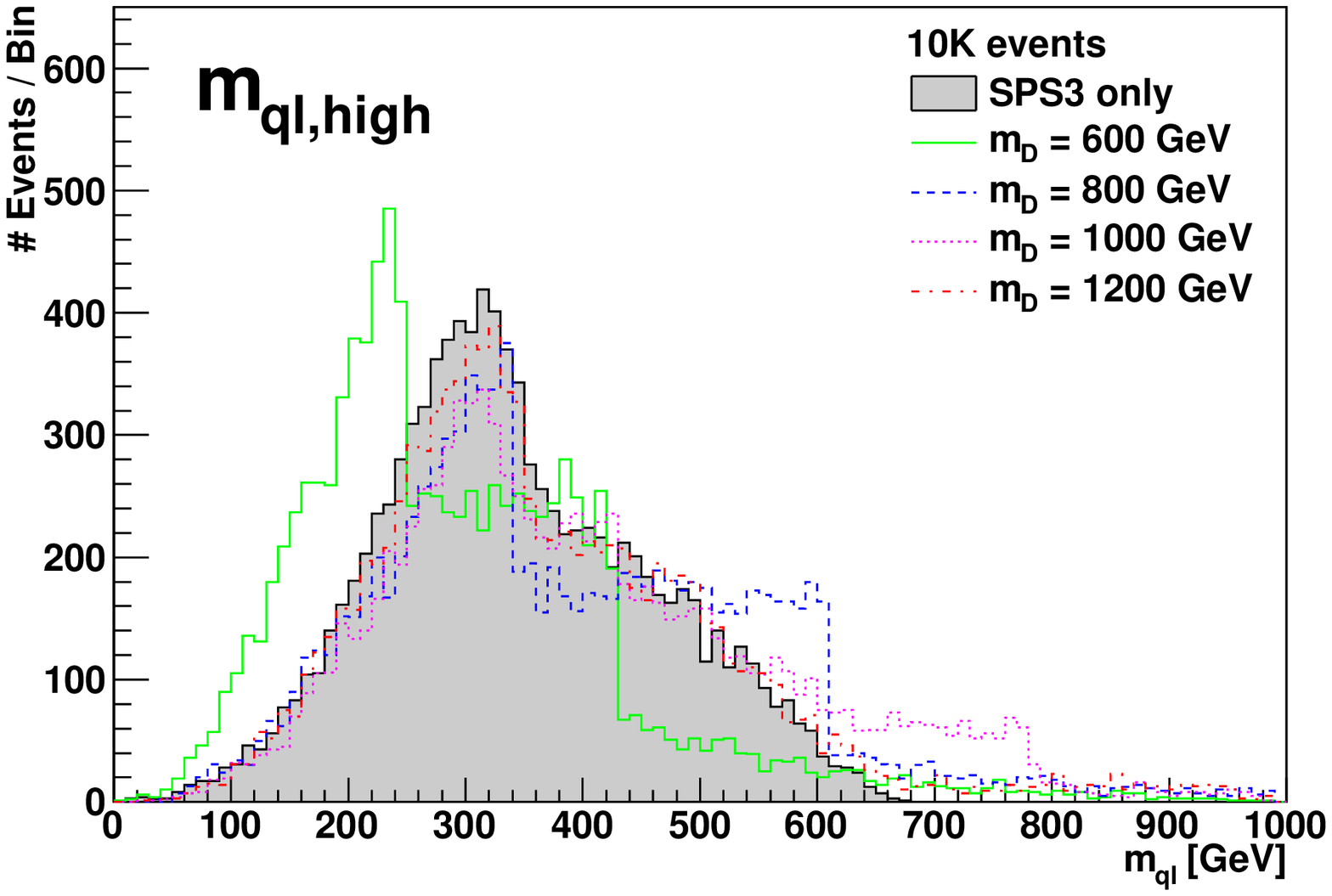} \\

      \vspace{-4mm}

      \includegraphics[width=0.45\textwidth]{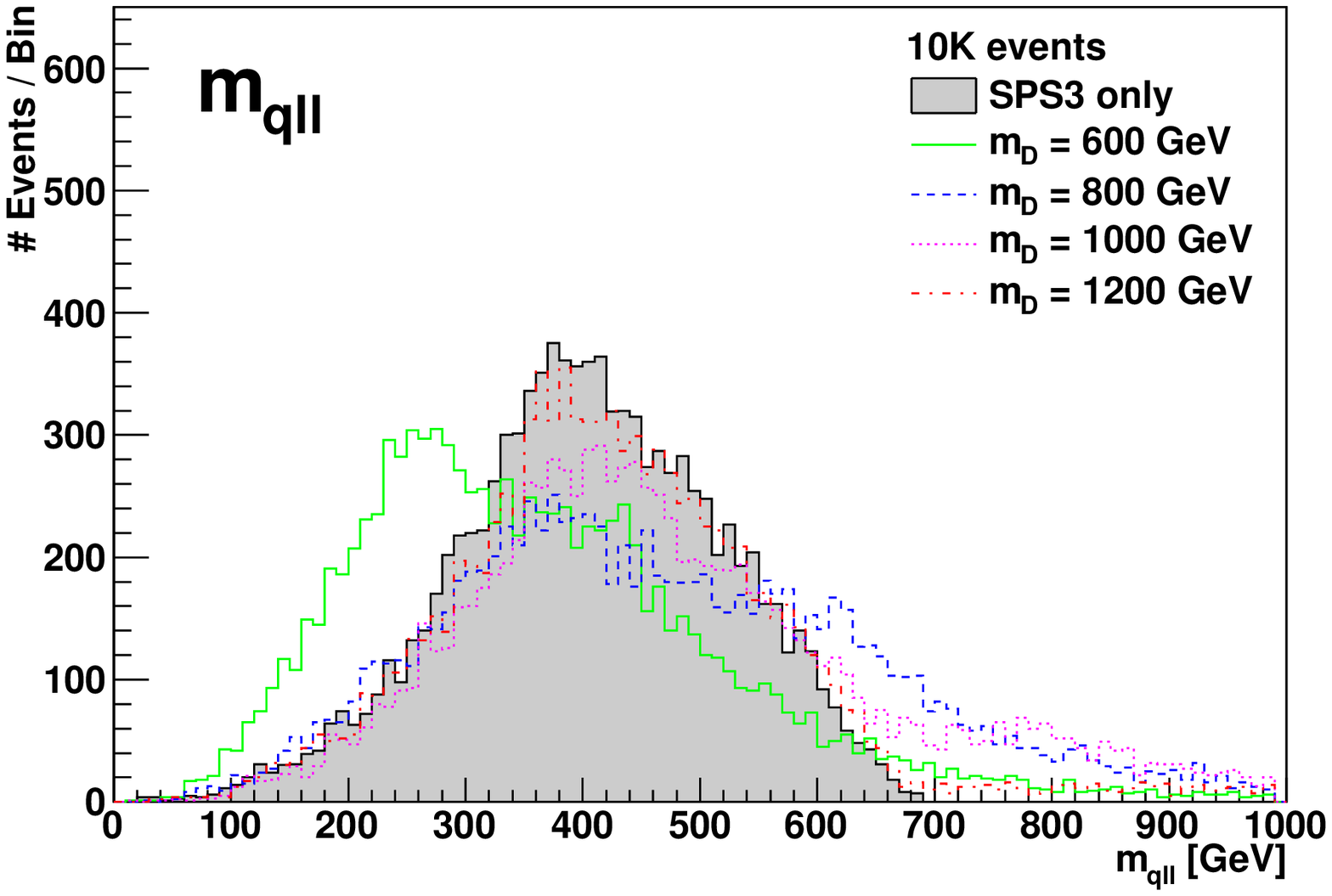} 
      \includegraphics[width=0.45\textwidth]{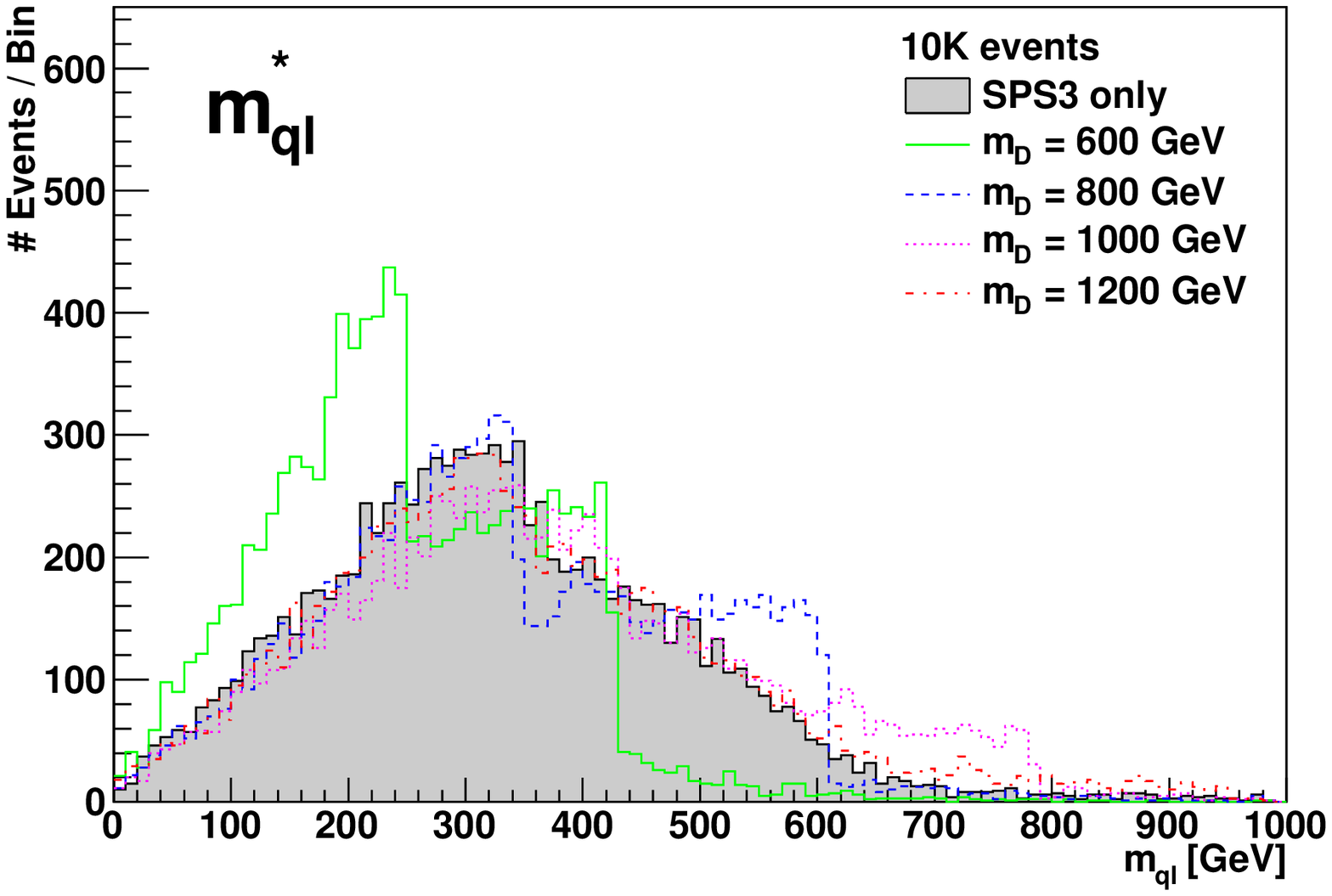} 
      \caption{Same as Fig.~\ref{fig:scan2}, but for SPS3.}
  \label{fig:scan3}
\end{figure}

\begin{figure}
    \centering
      \includegraphics[width=0.45\textwidth]{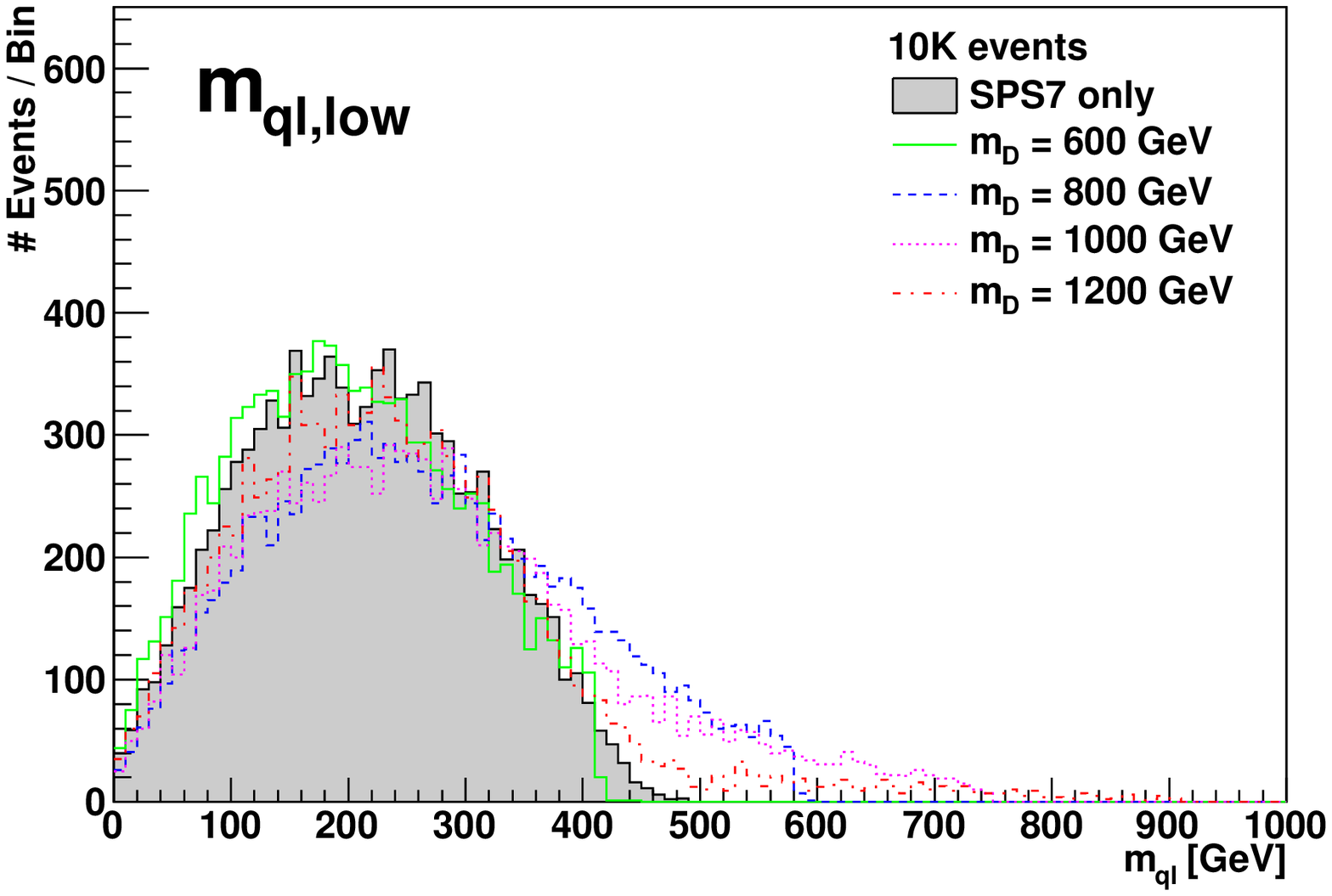} 
      \includegraphics[width=0.45\textwidth]{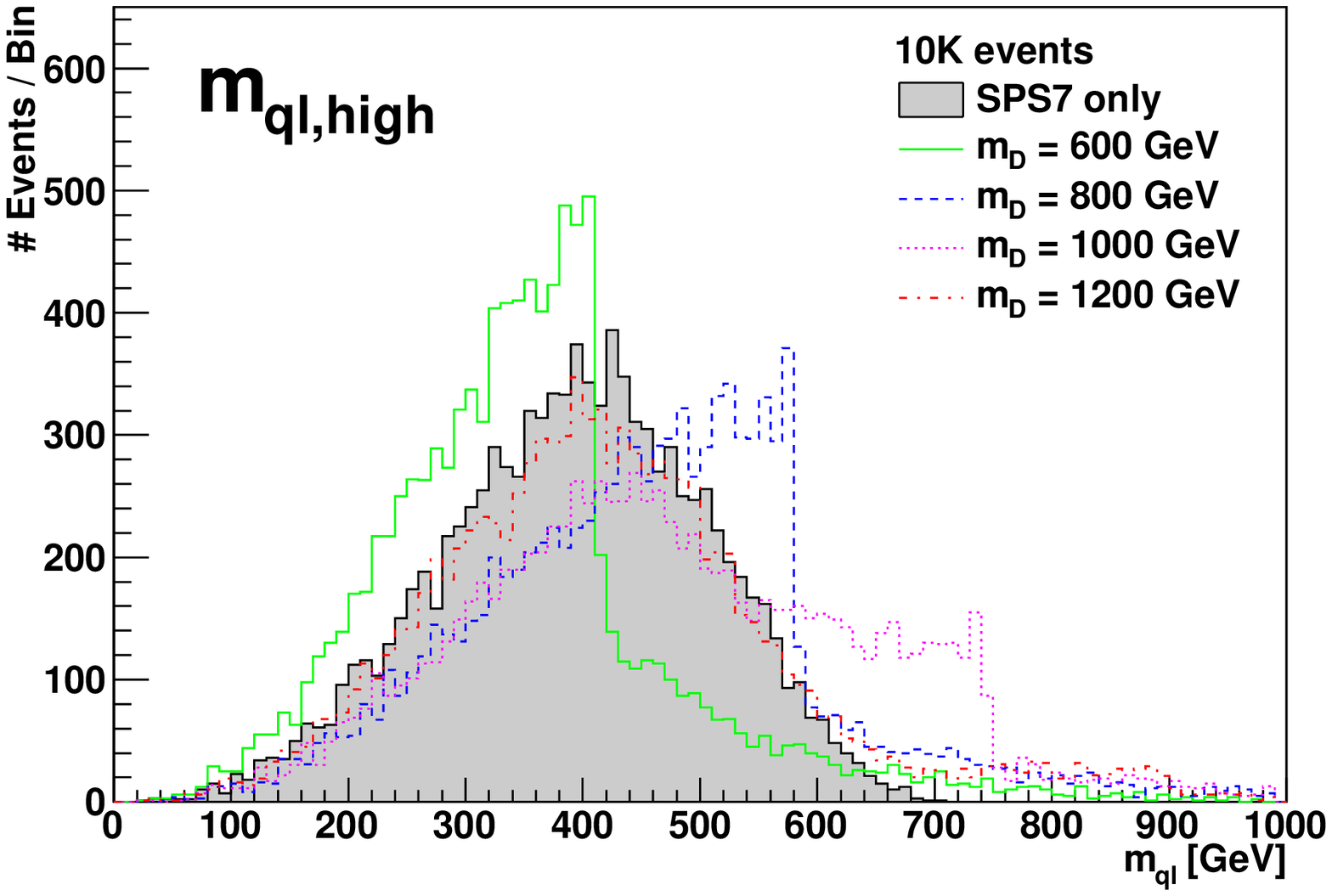} \\

      \vspace{-4mm}

      \includegraphics[width=0.45\textwidth]{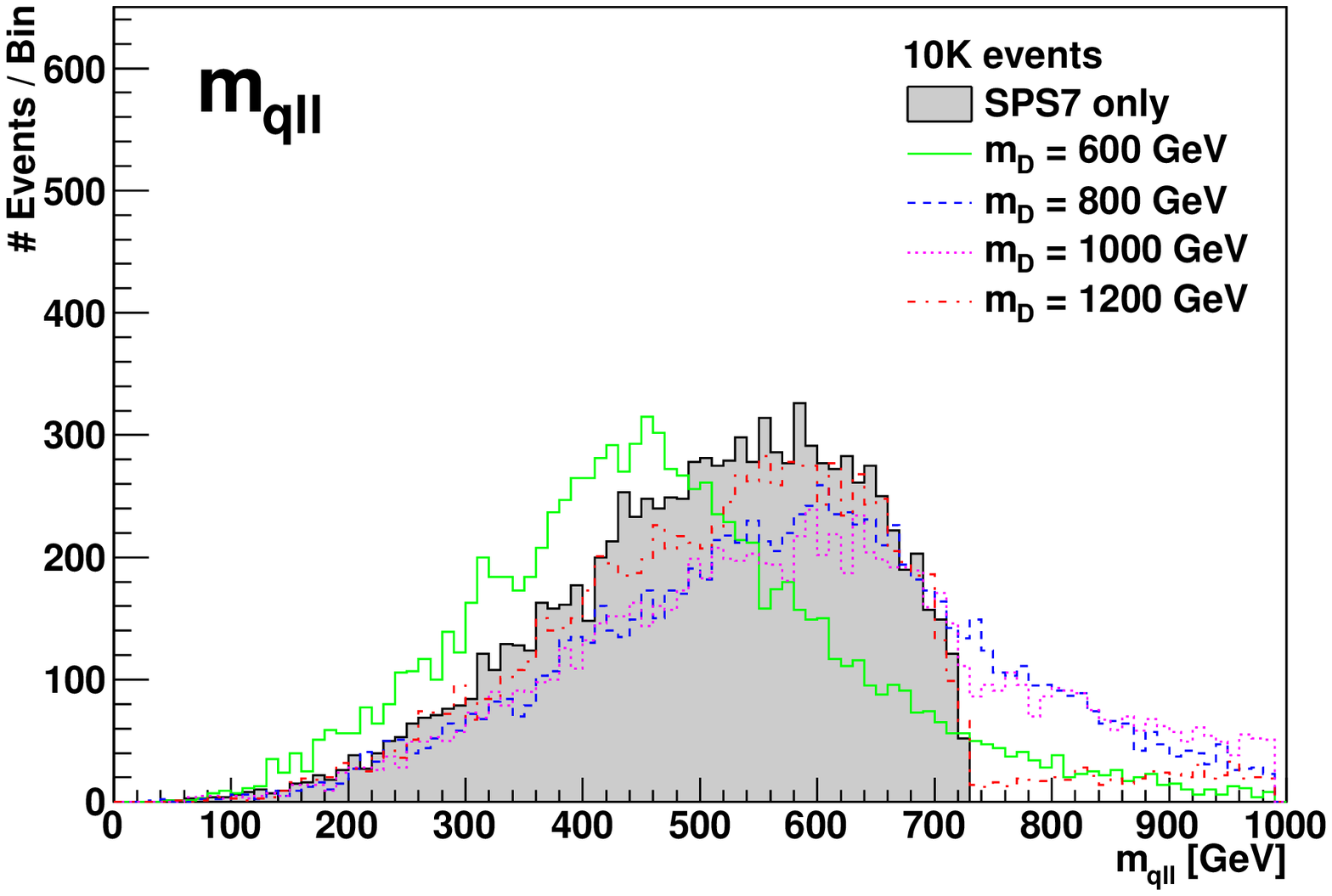} 
      \includegraphics[width=0.45\textwidth]{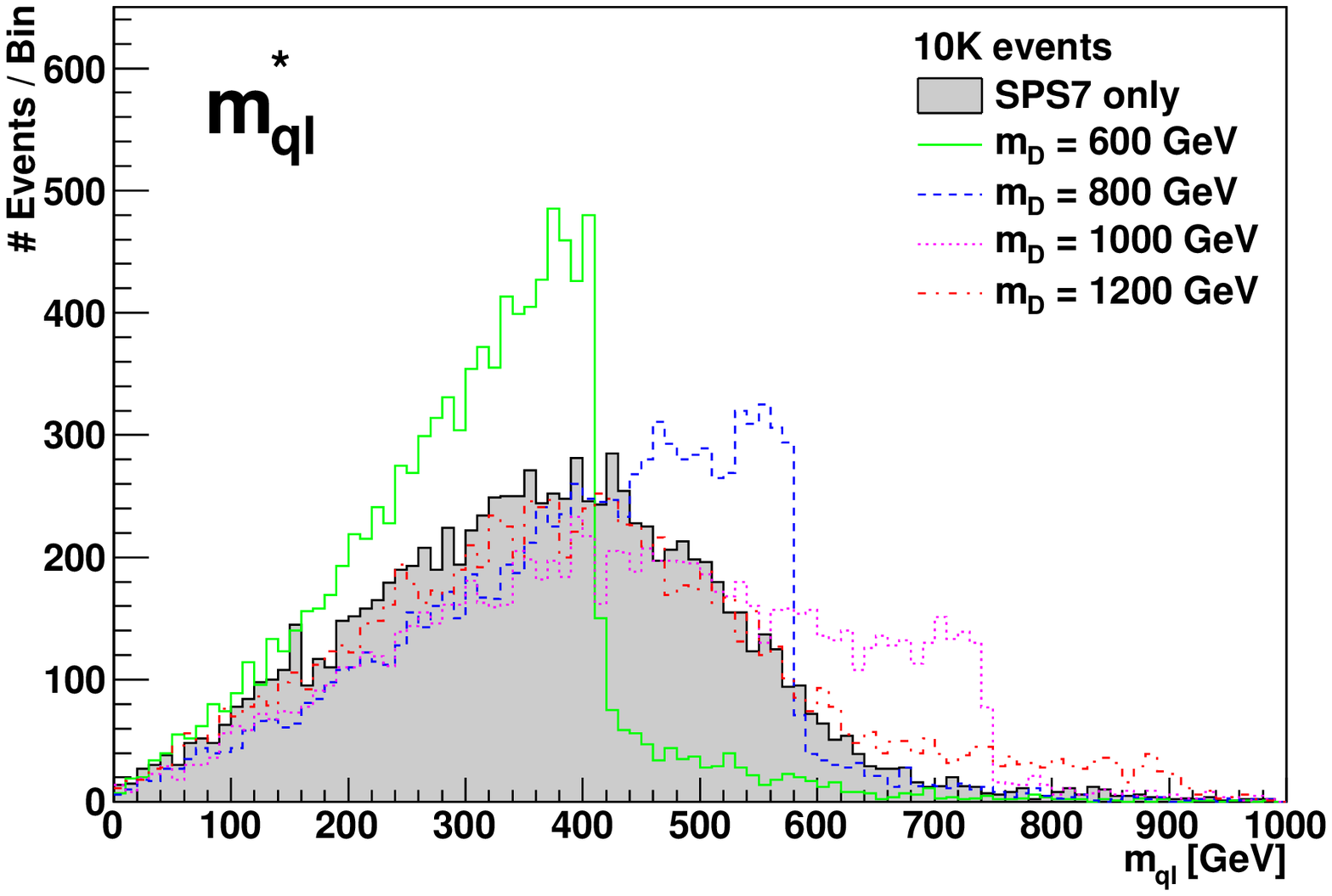} 
      \caption{Same as Figs.~\ref{fig:scan2},\ref{fig:scan3}, but for SPS7.} 
  \label{fig:scan4}
\end{figure}

In this appendix we collect the mass edge variables
$m_{ql,\text{low}}$, $m_{ql,\text{high}}$, $m_{qll}$, and $m_{ql}^*$
for varying SUSY spectra given by the Snowmass points 
SPS1b, SPS3, and SPS7, respectively in Fig.~\ref{fig:scan2},
Fig.~\ref{fig:scan3}, and Fig.~\ref{fig:scan4}. 



\end{document}